\documentclass[sigconf,nonacm]{acmart}
\title[A Comprehensive Analysis of Evolving Permission Usage in Android Apps]{A Comprehensive Analysis of Evolving Permission Usage in Android Apps: Trends, Threats, and Ecosystem Insights}

\author{Ali Alkinoon}
\orcid{0009-0002-8144-5302}
\affiliation{%
  \institution{University of Central Florida}
  \city{Orlando}
  \state{FL}
  \country{USA}
}
\email{mohaisen@ucf.edu}

\author{Trung Cuong Dang}
\orcid{0009-0002-5681-7223}
\affiliation{%
  \institution{University of Central Florida}
  \city{Orlando}
  \state{FL}
  \country{USA}
}\email{cuong.dang@ucf.edu}

\author{Ahod Alghuried}
\orcid{0009-0002-5681-7223}
\affiliation{%
  \institution{University of Central Florida}
  \city{Orlando}
  \state{FL}
  \country{USA}
}\email{AhodTareq.Alghuried@ucf.edu}

\author{Abdulaziz Alghamdi}
\orcid{0009-0008-8508-9281}
\affiliation{%
  \institution{University of Central Florida}
  \city{Orlando}
  \state{FL}
  \country{USA}
}\email{abdulaziz.alghamdi@ucf.edu}

\author{Soohyeon Choi}
\orcid{0009-0002-1252-2263}
\affiliation{%
  \institution{University of Central Florida}
  \city{Orlando}
  \state{FL}
  \country{USA}
}\email{Soohyeon.Choi@ucf.edu}

\author{Manar Mohaisen}
\orcid{0000-0002-7270-0933}
\affiliation{%
  \institution{Northeastern Illinois University}
  \city{Chicago}
  \state{IL}
  \country{USA}
}
\email{m-mohaisen@neiu.edu}

\author{An Wang}
\orcid{0000-0002-1701-9176}
\affiliation{%
  \institution{Case Western Reserve University}
  \city{Cleveland}
  \state{OH}
  \country{USA}
}
\email{axw474@case.edu}

\author{Saeed Salem}
\orcid{0000-0001-6478-4674}
\affiliation{%
  \institution{Qatar University}
  \city{Doha}
  \country{Qatar}
}
\email{saeed.salem@qu.edu.qa}

\author{David Mohaisen}
\orcid{0000-0003-3227-2505}
\affiliation{%
  \institution{University of Central Florida}
  \city{Orlando}
  \state{FL}
  \country{USA}
}
\email{mohaisen@ucf.edu}

\begin{abstract}
The proper use of Android app permissions is crucial to the success and security of these apps. Users must agree to permission requests when installing or running their apps. Despite official Android platform documentation on proper permission usage, there are still many cases of permission abuse. This study provides a comprehensive analysis of the Android permission landscape, highlighting trends and patterns in permission requests across various applications from the Google Play Store. By distinguishing between benign and malicious applications, we uncover developers' evolving strategies, with malicious apps increasingly requesting fewer permissions to evade detection, while benign apps request more to enhance functionality. In addition to examining permission trends across years and app features such as advertisements, in-app purchases, content ratings, and app sizes, we leverage association rule mining using the FP-Growth algorithm. This allows us to uncover frequent permission combinations across the entire dataset, specific years, and 16 app genres. The analysis reveals significant differences in permission usage patterns, providing a deeper understanding of co-occurring permissions and their implications for user privacy and app functionality. By categorizing permissions into high-level semantic groups and examining their application across distinct app categories, this study offers a structured approach to analyzing the dynamics within the Android ecosystem. The findings emphasize the importance of continuous monitoring, user education, and regulatory oversight to address permission misuse effectively.
\end{abstract}

\keywords{Android, Permissions, Privacy, Measurement}

\newtheorem{takeaway}{Takeaway}
\usepackage{amsthm}
\usepackage[inline]{enumitem}

\usepackage{xspace}
\usepackage{pifont}
\newcommand{\BfPara}[1]{\noindent\textbf{#1.}\xspace}
\newcommand{\etal}{\textit{et al.}\xspace}

\usepackage{makecell}
\newcommand{\xl}[1]{\Xhline{#1\arrayrulewidth}}

\usepackage{multirow}
\usepackage{xcolor}

\usepackage[framemethod=TikZ]{mdframed}
\definecolor{boxcolor}{RGB}{240, 248, 255}
\newmdenv[
  backgroundcolor=boxcolor,
  linewidth=0pt,
  innerleftmargin=10pt,
  innerrightmargin=10pt,
  innertopmargin=10pt,
  innerbottommargin=10pt
]{takeawaybox}

\usepackage{pgfplots}
\usepackage{amsmath}

\usepackage{xcolor}

\begin{document}
\maketitle

\section{Introduction}

Android holds a market share of 72.2\%, making it the most widely used mobile operating system globally, powering billions of devices from smartphones to tablets and wearables~\cite{SenanayakeKA21}. Its open-source nature and comprehensive app ecosystem have driven its popularity. Android's versatility and adaptability have made it a cornerstone of mobile technology, fostering innovation and accessibility across various applications. As of 2024, the Google Play Store offers an extensive collection of about 2.4 million mobile apps~\cite{number-of-google-apps}, catering to the diverse preferences of Android users. This prolific app landscape serves a user base of 3.6 billion device owners~\cite{num-of-users}.

The permission system is central to Android apps' functionality and security~\cite{ReardonFWOVE19}, which governs access to sensitive data and essential device functions. Permissions regulate access to personal data, such as contacts, messages, location, and hardware features, like the camera and microphone~\cite{0001BDMOW16}. With its ample features, this system empowers users to control the data and functionalities that apps can access, thereby protecting their privacy and enhancing security~\cite{WijesekeraBHEWB15}. The Android permission system operates on the principle of least privilege, meaning that apps should only request permissions necessary for their core functionality~\cite{LiDLDG21,WeiGNF12}. This minimizes abuse potential and reduces security breach risk. When an app is installed, it requests permissions specified in its manifest file. Users must grant these permissions for the app to function correctly~\cite{MalviyaTLXSJ23}.

The study of app permissions within the Android ecosystem is paramount due to several critical factors. Firstly, permissions are crucial gatekeepers to sensitive data and essential device functionalities~\cite{EnckOMC11}. Improperly managing these permissions, such as over-requests or application misuses, introduces substantial security and privacy risks~\cite{FeltCHSW11}. User personal information, including contacts, messages, and location data, can be susceptible to unauthorized access and malicious exploitation. This underscores the importance of scrutinizing how permissions are requested and used. This will mitigate potential security threats and enhance the Android platform's integrity. Permission management has significant implications for both developers and users. From a user perspective, understanding and controlling app permissions is crucial to maintaining privacy and trust. Users are more likely to engage with and recommend apps that are transparent about their data usage and request only the necessary permissions.


Understanding the {\em status quo} of the Android permission system and its use across various applications through statistical analysis, correlational understanding, and contrast is critical. However, this topic is not new, and several studies have examined this question (as highlighted in~\autoref{sec:relatedwork}). Nonetheless, we believe this pursuit is still important for the following reasons. First, we examine a new and fresh set of applications that reflect an up-to-date view of the permissions landscape, capturing a more accurate characterization than the dated studies. Second, the Android permission system evolves and so do the permission usages. We hope to provide a current perspective by analyzing recently updated applications. Third, given the evolving legislative and regulatory mandates, trends in the use of Android permissions are likely changing. We hope to shed light on the most up-to-date view of those permission trends by capturing the policy evolution. Finally, a comprehensive, up-to-date understanding of the current landscape remains limited.

\BfPara{Contributions}In this paper, we make the following contributions:

\begin{enumerate}
    \item {\bf Longitudinal Permission Analysis.} We conduct a multi-year exploration of Android permission usage trends from 2019 to 2023, comparing how benign and malicious apps request permissions over time. This temporal period allows us to observe shifts in behavior, such as malicious apps minimizing permission use to evade detection.

    \item {\bf Genre-Based Comparative Study.} We examine permission usage across 16 app genres  (e.g., {\em Finance}, {\em Education}, {\em Communication}), offering one of the broadest category-level analyses to date. This genre-specific breakdown reveals nuanced permission patterns that are often obscured in aggregate analyses.

    \item {\bf Structured Semantic Categorization.} We introduce and apply a consistent, interpretable taxonomy of high-level semantic permission groups (e.g., Location and GPS, Network Connectivity). This categorization enables clearer cross-genre and cross-year comparisons, supporting more interpretable analysis than ad-hoc or inconsistent groupings in prior work.

    \item {\bf Association Rule Analysis} Leveraging the FP-Growth algorithm, we identify frequent permission combinations and patterns of co-occurrence across the entire dataset. Our analysis spans the full dataset, individual years, and specific app genres, uncovering meaningful trends in benign and malicious apps.

    \item {\bf Multi-Dimensional Feature Comparison.} We analyze additional app metadata, such as in-app purchases, content ratings, ad presence, app size, and user ratings, to study how these features correlate with permission behavior and potential risk. This holistic perspective helps uncover broader ecosystem patterns that affect both developers and users.

\end{enumerate}

\BfPara{Organization} This paper is structured as follows: we begin by reviewing significant related work in~\autoref{sec:relatedwork}, which sets the foundation for our research. Next, we provide a background of the Android permission domain in~\autoref{sec:background}, outlining the key concepts and structures that inform our study. In~\autoref{sec:Methodology}, we detail the data collection and analysis methodology, followed by a thorough presentation of our analysis results in~\autoref{sec:AnalysisResults}. These findings are comprehensively discussed in~\autoref{sec:Discussion}, highlighting the key takeaways. Finally, we address the study's limitations and offer recommendations in~\autoref{sec:Limitations}, and provide concluding remarks and suggestions for future research in~\autoref{sec:Conclusion}.


\section{Related Work} \label{sec:relatedwork}

Research on the Android permission system has evolved significantly, addressing various aspects, including system design, user comprehension, and security implications. Numerous studies have examined permission usage patterns, overprivilege detection, user perception, and malware detection strategies, providing valuable insights into Android's permission framework. Table~\ref{tab:related_work} summarizes key contributions from related work, including methodologies, features examined, and their limitations. Earlier works often focused on limited datasets, specific methodologies, or narrow perspectives (e.g., overprivilege detection or user perception), our work, in contrast, builds on and extends these efforts. Specifically, we analyze permission usage across app categories over a five-year period (2019--2023), incorporating additional features, such as advertisements and in-app purchases. This holistic approach enables a broader understanding of permission trends, their security implications, and their relationship with app functionalities.

\BfPara{Android Permissions System Overview} 
 
The Android permissions system has been the subject of extensive research focusing on its limitations, user understanding, security mechanisms, and system design. Studies have explored permission usage patterns and granularity~\cite{DiamantarisPMIP19, TaylorM16, DawoudB19, YangBRGI12}, while others emphasized enhancements to refine how permissions are managed~\cite{SellwoodC13, AuZHL12, QuRZCZC14, ChenJDDMMWRS13}. More recent works introduced dynamic and context-aware permission models to address evolving privacy challenges~\cite{WangSCL22, CaoXPLTA21, TuncayDGG18}. Barrera~\etal\cite{BarreraKOS10} used the Self-Organizing Map (SOM) algorithm to highlight permission usage patterns, identifying areas for refinement, while Almomani~\etal\cite{AlmomaniK20} provided an overview of Android's evolving permission framework.

\BfPara{Permission Optimization and Overprivilege Detection}
Overprivileged permissions and minimizing unnecessary requests are key issues in Android security research. Xiao~\etal\cite{XiaoCHFX20} introduced MPDroid, which combines static analysis and collaborative filtering to tackle overprivilege. Similarly, Johnson~\etal\cite{JohnsonWGS12} mapped Android API calls to required permissions, automating app downloads, and analyzing permission accuracy. Our work builds on this by categorizing permissions into semantic groups and comparing their usage with app features like ads and in-app purchases.

\BfPara{User Perception and Risk Signals}
Several works examine the link between app permissions and user perception. Sarma~\etal\cite{SarmaLGPNM12} integrated risk signals into permission warnings, and Felt~\etal\cite{FeltHEHCW12} identified user comprehension challenges. While these studies provide insights into user interaction, our research extends by exploring permission use in connection with app features and its effect on user privacy.

\BfPara{Longitudinal Studies on Android Permissions} Longitudinal research on permission systems helps to reveal trends and security risks over time. Wei~\etal\cite{WeiGNF12} conducted such a study, observing an increase in dangerous permissions over the years. Zhauniarovich~\etal\cite{ZhauniarovichG16} analyzed the transition to runtime permissions introduced in Android 6.0. Our research builds on these findings by examining how permission usage varies across app categories and changes over time. Specifically, we analyze the frequency and type of permissions requested within different app categories (e.g., {\em gaming}, {\em finance}, {\em education}) and observe shifts in these patterns over the years. This approach provides a more granular view of permission trends, highlighting category-specific behaviors and evolving practices in permission requests.

\BfPara{Permissions and App Features}
The relationship between permissions and app features, including ads, in-app purchases, and app trustworthiness has also been explored. Wang~\etal\cite{WangWTZW20} used natural language processing to study how permissions influence user trust, while Scoccia~\etal\cite{ScocciaPPMK19} examined how developers handle permission-related issues. Our research provides a more comprehensive comparison of permission usage and app features, offering insights into their role in app functionality and user privacy.

\BfPara{Permission Analysis for Security and Malware Detection}
Security research has extensively used permission analysis to enhance Android malware detection strategies. Li~\etal\cite{LiJYX23} and Guyton~\etal\cite{GuytonLWK22} both optimized feature selection by analyzing permissions, intents, and API calls, while Rathore~\etal\cite{RathoreSRS20} developed a malware detection system that strongly relies on permission data. Additionally, Mohaisen~\etal\cite{mohaisen2015amal} introduced AMAL, a behavior-based automated malware classification system that complements permission-centric approaches by examining static and dynamic behaviors at scale. Kang~\etal\cite{kang2015detecting} further advanced detection accuracy by incorporating creator information such as certificate serial numbers into static analysis pipelines for classification and attribution. Beyond Android, Alasmary~\etal\cite{alasmary2019analyzing} proposed a graph-based approach for detecting emerging malware in the Internet of Things (IoT), showing that metadata, structure, and behavioral context are vital for robust detection across platforms. While these studies focus mainly on security, our work broadens the scope by analyzing permissions not only in terms of security threats but also by exploring their broader implications on user privacy and overall app behavior.

\begin{table*}
\centering
\caption{Comparison of Related Work on Android Permissions Research.}\label{tab:related_work}\vspace{-3mm}
\scalebox{0.85}{
\begin{tabular}{lccllll}
\xl{2}
\textbf{Author} & \textbf{Year} & \textbf{Samples} & \textbf{Method} & \textbf{Apps} & \textbf{Features} & \textbf{Limits} \\
\xl{1}
Wei \etal~\cite{WeiGNF12}              & 2009--2011 & 237      & Longitudinal study              & Perm. changes      & Dangerous, pre-installed perms   & No comp. w/ features   \\
Barrera \etal~\cite{BarreraKOS10}      & 2010--2011 & 1,100    & SOM clustering                  & Perm. analysis     & Use patterns, granularity        & Few cats., early study \\
Johnson \etal~\cite{JohnsonWGS12}      & 2012       & 141,000  & API map + auto DL               & Perm. accuracy     & Misuse detection                 & No feat. mapping       \\
Sarma \etal~\cite{SarmaLGPNM12}        & 2012       & 158,062  & Risk signal fusion              & Warn. decisions    & Risks vs. benefits               & Shallow analysis       \\
Felt \etal~\cite{FeltHEHCW12}          & 2012       & 333      & User studies                    & Perm. effectiveness& Attn., understanding             & No feat. links         \\
Zhauniarovich \etal~\cite{ZhauniarovichG16} & 2016 & --       & Runtime analysis                & Perm. system       & Dynamic, structural mgmt.        & No comp. eval          \\
Guyton \etal~\cite{GuytonLWK22}        & 2018       & 119K     & Sec. model opt.                 & Malware detect.    & Perms, intents, APIs             & No privacy impact      \\
Wang \etal~\cite{WangWTZW20}           & 2019       & 20K      & NLP on reviews                  & User feedback      & Trust, user perception           & Lacks deep analysis    \\
Xiao \etal~\cite{XiaoCHFX20}           & 2020       & 16,343   & Static + CF                     & Overpriv. detect.  & Min. necessary perms             & Narrow scope           \\
Rathore \etal~\cite{RathoreSRS20}      & 2021       & 11,281   & Perm.-based ML                  & Malware detect.    & High-risk perms                  & No feat. eval          \\
Scoccia \etal~\cite{ScocciaPPMK19}     & --         & 574      & Exploratory                     & Perm. mgmt.        & Issue fixing, practices          & No func. insight       \\
Li \etal~\cite{LiJYX23}                & --         & 814      & Static + Apriori                & Dev guidance       & Perm. relationships              & No behavior links      \\
Almomani \etal~\cite{AlmomaniK20}      & --         & --       & SOM clustering                  & Perm. framework    & Dev risks, vuln. focus           & No cat. detail         \\
\xl{1}
\textbf{This work}                     & 2019--2023 & 5,028    & Category + comp. analysis       & Perm. profiling    & Genres, semantics, ads, IAPs     & Partial feat. coverage \\
\xl{2}
\end{tabular}
}
\end{table*}

\section{Background}\label{sec:background}
Understanding the context of Android permissions is crucial for analyzing their impact and usage within the ecosystem. This background section provides an overview of essential components in the Android permission domain. We begin by discussing the Google Play Store, the primary distribution platform for Android apps, followed by an exploration of Android Application Packages (APKs), which serve as the core unit for app delivery and installation. Finally, we delve into the intricacies of Android permissions, examining how they govern app behavior and user privacy.

\begin{figure}[htb]
    \centering
    \includegraphics[width=0.99\linewidth]{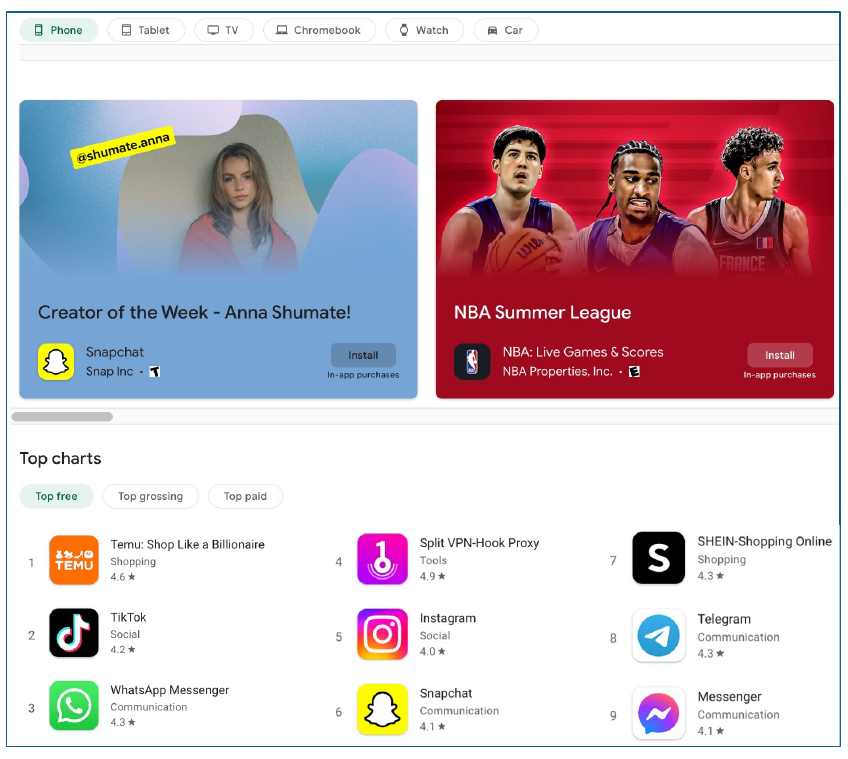}
    \caption{A screenshot of the Google Play Store (web interface) highlighting top charts, platforms, and categories.}
    \label{fig:appstore}\vspace{-5mm}
\end{figure}

\BfPara{Google Play Store} \label{sec:GPstore}
Google Play Store~\cite{GoogleStore} is an online platform and digital distribution service that serves as the official app store for Android devices. It offers a centralized hub for discovering, purchasing, and managing apps, games, movies, music, books, and other digital content. Users can access free and paid content through the Play Store app or its web interface, as illustrated in Figure~\ref{fig:appstore}.

\BfPara{Understanding an APK}An Android Application Package (APK) is the standard file format for distributing and installing apps on Android. It contains all the necessary components such as code, libraries, assets, and a manifest file needed to seamlessly run the app~\cite{DongLDLLLXCWZ18}. The manifest file, {\tt AndroidManifest.xml}, provides crucial information like the package name, permissions, and hardware requirements, allowing proper system execution and robust security enforcement. The assets folder contains uncompiled data, such as text, images, and audio files, accessible during run-time. The {\tt resources.arsc} file stores essential UI resources, while the {\tt classes.dex} file contains compiled Java bytecode for the app. Finally, the META-INF directory houses metadata and signature files, ensuring the overall integrity of the APK by preventing tampering.

\BfPara{Permissions in Android Platform}App permissions in  Android are security measures designed to protect user data and ensure privacy by regulating what actions an app can perform and what information it can access. When an app requests specific permission, it seeks authorization to access certain features or data on the user's device, including personal information, system resources, or device hardware. Users are prompted to grant or deny these permissions during the app's installation or while using the app.

Permissions in Android are officially categorized by Google into two types: normal and dangerous, as defined in the Android Developer Documentation~\cite{xxx}. The normal permissions cover less sensitive operations, such as internet access. These are automatically granted at the installation time because they pose minimal risk to the user's privacy or device security. The dangerous permissions, on the other hand, involve access to more sensitive data and require explicit consent. Examples include permissions for accessing the user's location, contacts, and camera. Beginning with Android 6.0 ({\tt Marshmallow}), apps must request dangerous permissions at runtime, giving users more control and transparency over their data~\cite{MalviyaLKTSJ22}. For instance, a social media app might request access to the camera for photo uploads, while a navigation app would require location access to provide accurate directions. By requiring run-time approval, Android ensures that apps cannot access sensitive data without the user's knowledge and consent.

\begin{figure}[t]
    \centering
    \includegraphics[width=0.99\linewidth]{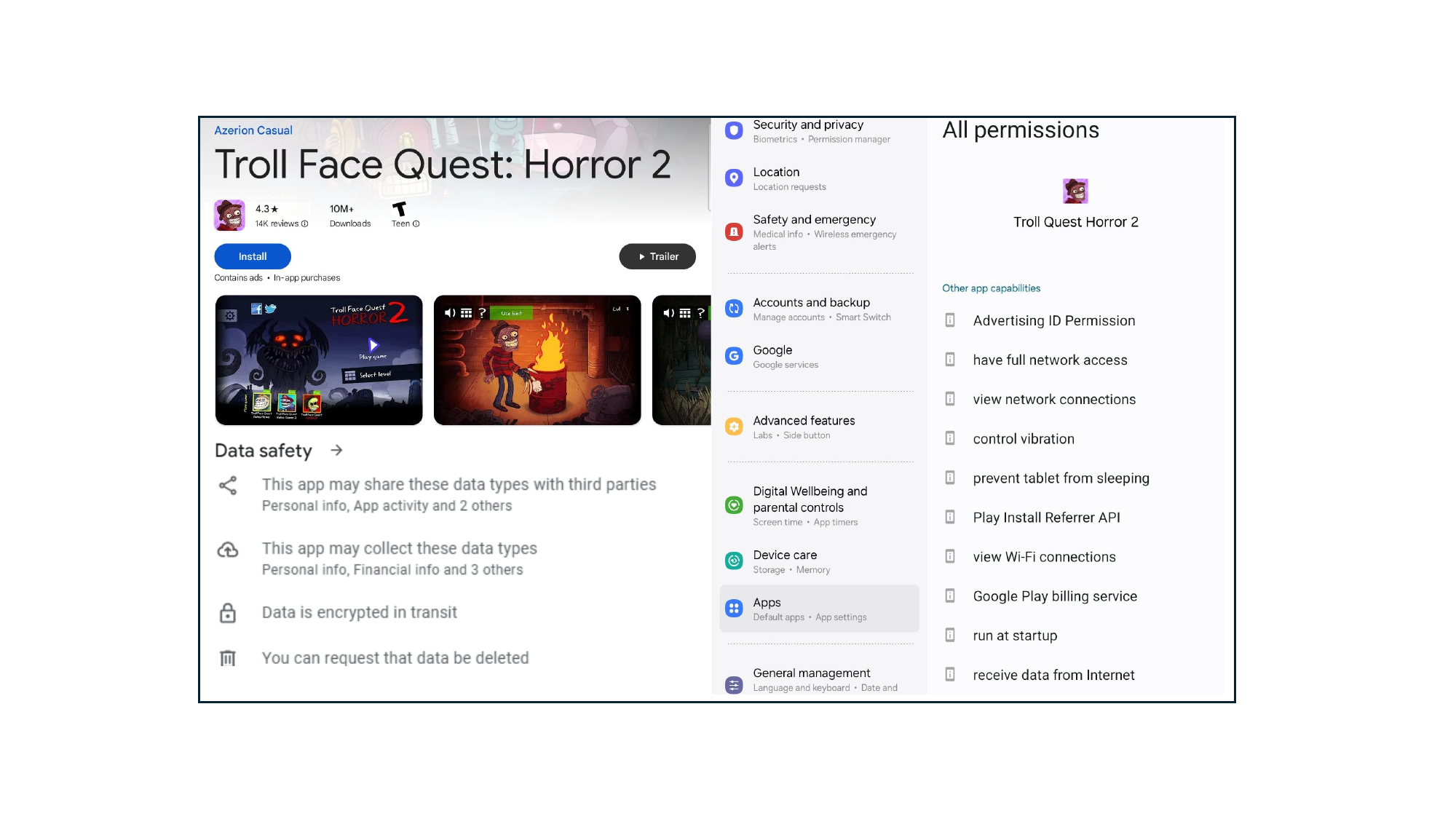}
    \caption{A screenshot of an app within our dataset displaying its associated requested permissions.}
    \label{fig:AppPermissions}\vspace{-5mm}
\end{figure}

Location access permissions, such as {\tt ACCESS\_FINE\_LOCATION}, allow an app to retrieve precise user location data using GPS and network-based sources. In contrast, {\tt ACCESS\_COARSE\_LOCATION} permits access to approximate location information derived from Wi-Fi and cell towers. Similarly, permissions for device hardware, such as {\tt CAMERA} and {\tt RECORD\_AUDIO}, enable the app to use the device's camera and microphone to capture photos, record videos, or capture audio. Storage permissions, like {\tt READ\_EXTERNAL\_STORAGE} and {\tt WRITE\_EXTERNAL\_STORAGE}, grant the app access to external storage, including photos, videos, and other files. These permissions are often presented at a higher semantic level in user-facing interfaces, abstracting technical permission names into descriptions of functionality. For example, {\tt ACCESS\_NETWORK\_STATE} may be displayed as ``have full network access,'' as demonstrated in Figure~\ref{fig:AppPermissions}. The snapshot showcases an example app from our dataset, visually highlighting its requested permissions and their representations alongside key app metadata, such as data safety and usage details.

\section{Methodology}\label{sec:Methodology}
\begin{figure*}[htb]
    \centering
    \includegraphics[width=\linewidth]{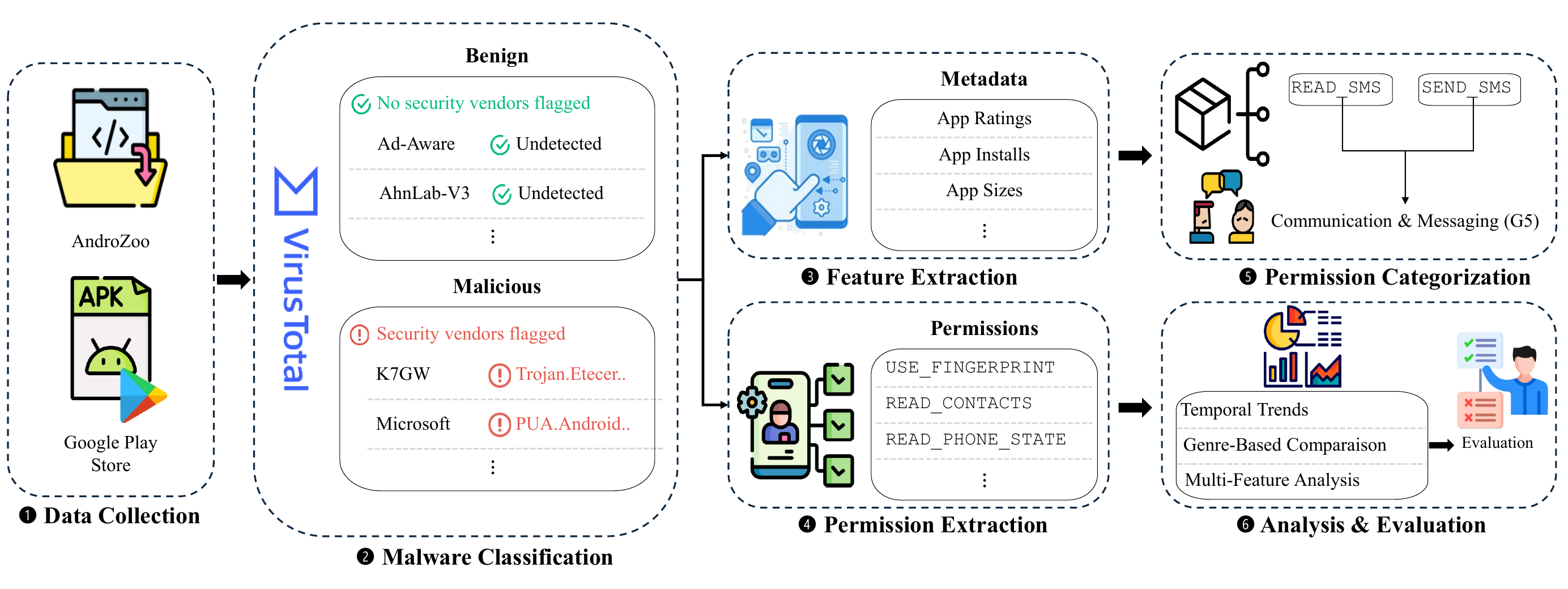}\vspace{-3mm}
    \caption{Overview of the analysis pipeline used in this study. The process includes data collection from AndroZoo and the Google Play Store, malware classification using VirusTotal~\cite{MohaisenA14}, feature and permission extraction, semantic permission categorization, and multi-dimensional analysis involving temporal trends, genre-based comparisons, and metadata correlation.}
    \label{fig:Pipeline}\vspace{-3mm}
\end{figure*}

We adopted a structured method for collecting, labeling, and analyzing Android application data, with a particular emphasis on permission usage and its broader implications. The process is organized into several core stages: data collection, malware classification, feature extraction, permission extraction, permission categorization, and comprehensive analysis. Each stage is designed to maintain dataset integrity and support reliable, reproducible insights. An overview of the full pipeline is illustrated in Figure~\ref{fig:Pipeline}, providing a visual summary of the workflow used throughout this study.

\BfPara{\ding{182} Data Collection}\label{sec:DataCollection} We obtained our dataset from AndroZoo~\cite{AllixBKT16}, a comprehensive source for Android apps' data. AndroZoo's data collection process prioritizes two primary features to enhance dataset robustness and versatility. First, it spans a significant temporal range, ensuring the inclusion of apps from various periods, which supports comprehensive analyses across different studies. Second, it emphasizes sourcing apps from reputable and well-established markets, particularly the Google Play Store, to ensure the integrity and credibility of the collected apps. To achieve this, we implemented a verification step during preprocessing in which each app's unique AppID was cross-referenced against current listings on the Google Play Store. This ensured that our final dataset includes only apps that are actively available on the platform at the time of analysis. Initially, we collected 7,000 apps cross-validated with the Google Play Store to ensure their presence on the official Android market. This validation process resulted in a final dataset of 5,028 apps, comprising 4,465 benign and 563 malicious apps, spanning five years from 2019 to 2023.

\BfPara{\ding{183} Maliciousness Classification} To classify Android applications in our dataset as benign or malicious, we relied on VirusTotal~\cite{VirusTotal}, a widely used platform that analyzes APK files using over 70 antivirus engines\cite{MohaisenA14}. We first checked whether an app had an existing VirusTotal scan result from the AndroZoo repository. If no scan result was available or if the scan was outdated, we manually submitted or re-scanned the APK on VirusTotal to ensure up-to-date results. We adopted a sensitive labeling policy: an app was labeled as malicious if at least one antivirus engine flagged it, and labeled as benign if no engines flagged it. This ensured consistency across the dataset while capturing a broad range of potentially harmful behaviors. This dual approach, leveraging existing annotations and performing fresh scans, helped enhance the reliability of our labels and reduce ambiguity stemming from stale metadata, a common practice in the prior work~\cite{WangMCC15, WangCCM18, ZapzalkaSM25, AbusnainaASAJSM24, ChoiAAASWCNAM22, AbusnainaAAAJSN22, AbusnainaAAAJNM22, AlasmaryAJAANM20, AlasmaryKAPCAAN19, ShenVMKZ19}. Previous work has examined the reliability and consistency of antivirus labels across vendors, highlighting discrepancies and challenges in using them as ground truth for malware classification~\cite{MohaisenA14}.

\BfPara{\ding{184} Feature Extraction}\label{sec:FeatureExtraction} After assembling our dataset, we extracted and consolidated a comprehensive set of metadata features for each APK using a combination of the AndroZoo platform and a Google Play Store metadata scraper~\cite{GPscrapper}. These tools allowed us to systematically retrieve relevant app characteristics, including genre, ad-supported status, in-app purchases, content rating, app rating, install count, and APK size. These features are essential for analyzing trends in permission usage across different dimensions. Our analysis spans 16 distinct app genres, as detailed in~\autoref{tab:permissions_per_genre1-8} and~\autoref{tab:permissions_per_genre8-16}, providing a diverse and representative view of Android applications on the Play Store. This enriched metadata enabled more nuanced and comparative analyses of permission requests in relation to app functionality and user-facing traits.

\BfPara{\ding{185} Permission Extraction} \label{sec:PermissionExtraction}
The next step involved extracting permissions from each app. This process began with the decompilation of {\tt classes.dex} files from the APKs to obtain the Java source files that represent the applications. We systematically extracted the permissions using these source files and cataloged them in CSV format. This format included each app's package name and associated permissions, providing a structured approach to analyzing and assessing permission requests from various applications. From the 5,028 apps, we successfully extracted permissions from 4,136 benign apps and 343 malicious apps, resulting in 63,480 permissions. 

\autoref{tab:overall_stats} presents a detailed overview of the number of applications and the permissions extracted for each year, including the number of apps for which no permissions were found. In our initial analysis, a subset of applications returned the result ``No Permissions Found'' during the permission extraction phase. This outcome, based on static analysis of the decompiled source code using JADX and regular expression matching for permission references (e.g., {\tt android.permission.X}), indicates that these applications genuinely do not request any permissions. This breakdown includes 5,318 permissions from malicious apps and 58,162 permissions from benign apps, averaging 14 permissions per benign app and 16 per malicious app. A total of 321 unique permissions were identified across all categories, reflecting diverse usage and functionalities.

\begin{table}[]
\centering
\caption{Summary of apps count over years and types. }\label{tab:overall_stats}\vspace{-3mm}
{\begin{tabular}{c|c|ccc|ccc}
\xl{2}
\multirow{2}{*}{Year}  & \multirow{2}{*}{Apps} & \multicolumn{3}{c|}{Benign} & \multicolumn{3}{c}{Malicious} \\
\cline{3-8}
& & Total & Valid & Failed & Total & Valid & Failed \\
\xl{1}
2019  & 778        & 653          & 593               & 60          & 125             & 110               & 15         \\
2020  & 827        & 737          & 699               & 38          & 90              & 65                & 25         \\
2021  & 1,033      & 915          & 860               & 55          & 118             & 73                & 45         \\
2022  & 1,236      & 1,107        & 1,066             & 41          & 129             & 62                & 67         \\
2023  & 1,154      & 1,053        & 918               & 135         & 101             & 33                & 68         \\
\hline
\textbf{Total} & 5,028      & 4,465        & 4,136             & 329         & 563             & 343               & 220   \\
\xl{2}
\end{tabular}}\vspace{-5mm}
\end{table}

\BfPara{\ding{186} Permission Categorization} \label{sec:PermissionCategorization}
Given the extensive permissions in our dataset, we systematically categorized each permission into higher-level semantic categories. For instance, permissions such as {\tt INTERNET} and {\tt ACCESS\_NETWORK\_STATE} were grouped under the category of Network and Connectivity. This method of categorization was designed to provide a structured framework that enables a more coherent and meaningful analysis of permission usage. By organizing permissions in this way, we can better understand how different permissions are used across various apps, which in turn reveals patterns and insights into app behaviors and implications.

We grouped the 321 permissions by function, following the Android documentation. Using language models, we refined these groupings based on semantic similarities. This resulted in high-level categories: ``system and device management'' (G1, 92), ``network and connectivity'' (G2, 29), ``data access and storage'' (G3, 25), ``location and GPS'' (G4, 10), ``communication and messaging'' (G5, 28), ``media and camera'' (G6, 4), ``security and privacy'' (G7, 31), ``system UI and notification'' (G8, 16), ``app management and admin'' (G9, 50), and ``payment and transactions'' (G10, 36). The final categorization, which is illustrated in \autoref{tab:Permissions_Categories}, outlines these higher-level semantic groups along with the associated count of permissions within each category. This categorization is the foundation for our subsequent analysis, providing a clear and organized perspective on the Android permission landscape.

While the prior work has performed coarse semantic grouping of permissions (e.g., grouping based on functions like network or storage), our contribution lies in constructing and applying a comprehensive, consistent, and interpretable high-level semantic taxonomy that spans all permission types and supports longitudinal and category-specific analyses. We explicitly define 10 well-structured permission groups (e.g., G1: System \& Device Management, G4: Location \& GPS, etc.) and systematically apply this categorization in year- and genre-wise analyses. This structured framework enables clearer comparisons across app types and behaviors, which is often missing or inconsistently applied in earlier studies. For example, Xu \etal~\cite{XuTS2021} grouped permissions by function in their risk models, but do not apply a standardized taxonomy across all evaluation axes (e.g., time, category). Our work builds upon this direction with a consistent semantic scheme designed for interpretable analysis.

\begin{table}[t]
\centering
\caption{Breakdown of 321 Unique Permissions by Category}\label{tab:Permissions_Categories}\vspace{-3mm}
{\begin{tabular}{lr}
\xl{2}
\textbf{Category}                   & \textbf{Permissions} \\ \xl{1}
System and Device Management          & 92                         \\ 
Network and Connectivity              & 29                         \\ 
Data Access and Storage               & 25                         \\ 
Location and GPS                      & 10                         \\ 
Communication and Messaging           & 28                         \\ 
Media and Camera                      & 4                          \\ 
Security and Privacy                  & 31                         \\ 
System UI and Notifications           & 16                         \\ 
App Management and Administration     & 50                         \\ 
Payment and Transactions              & 36                         \\ 
\xl{1}
\textbf{Total}                        & 321                        \\ \xl{2}
\end{tabular}}\vspace{-3mm}
\end{table}

\BfPara{\ding{187} Analysis} \label{sec:Analysis}
We conducted a comprehensive analysis of Android permissions by identifying the top requested permissions across applications from multiple years (2019--2023) and within 16 distinct app genres, examining both the benign and malicious aspects for each year and genre. To deepen our understanding of permission usage patterns, we employed association rule mining with the FP-Growth algorithm \cite{FPGrowth}, which enabled us to uncover frequent permission combinations and highlight patterns of co-occurring permissions. Additionally, we examined the association between requested permissions and various app features, including genre, ad-supported status, in-app purchases, content rating, app rating, install base, and app size. This multifaceted analysis provided valuable insights into permission request trends, revealing key differences in permission behavior between benign and malicious apps, and highlighted specific patterns relevant to different app categories. Our findings contribute to a deeper understanding of permission usage dynamics and their implications for user privacy and security across diverse application types.

\section{Analysis Results}\label{sec:AnalysisResults}

This section presents a comprehensive analysis of permission requests in Android applications and their comparison.

\subsection{Top Requested Permissions}

\subsubsection{Top Requested Permissions by Year}

We analyzed the top permissions requested for benign and malicious apps from 2019 to 2023 to identify shifting trends in permission requests. The results revealed several key insights. \autoref{tab:permissions_per_year} provides detailed data on these trends. Across all years, the three most frequently requested permissions were {\tt ACCESS\_NETWORK\_STATE}, {\tt ACCESS\_FINE\_LOCATION}, and {\tt ACCESS\_COARSE\_LOCATION}, which fall under two dominant high-level categories: ``location and GPS'' (G4) and ``network and connectivity'' (G2). These categories reflect the core functionality of most Android apps, including location tracking and internet access.

Over time, malicious apps showed a consistent reduction in permission requests across all categories. For instance, {\tt ACCESS\_FINE\_\allowbreak LOCATION} dropped from 101 requests in 2019 to just 31 in 2023, and {\tt ACCESS\_NETWORK\_STATE} declined from 99 to 28. This trend suggests an evolving strategy to evade detection by minimizing sensitive permissions. Some permissions like {\tt CAMERA}, {\tt RECORD\_AUDIO}, and {\tt USE\_FINGERPRINT}, which fall under ``media and camera'' (G6) and ``security and privacy'' (G7), disappeared entirely from malicious apps by 2023. In contrast, benign apps generally maintained or increased use of certain sensitive permissions. Location-related permissions (G4) peaked in 2022, and network-related permissions (G2) remained consistently high. However, a few permissions, such as {\tt WRITE\_EXTERNAL\_\allowbreak STORAGE} and {\tt CAMERA}, showed a decline over time. Notably, permissions like {\tt STATUS\_BAR\_SERVICE} and {\tt MEDIA\_CONTENT\_\allowbreak CONTROL}, which fall under ``system UI and notification'' (G8) and ``media and camera'' (G6), respectively, were used exclusively by benign apps, highlighting distinct usage patterns between benign and malicious behaviors.

\begin{table}
\centering
\caption{The top requested permissions for benign (B) and malicious (M) over the years from 2019 to 2023.}
\label{tab:permissions_per_year}\vspace{-3mm}
\scalebox{0.68}{
\begin{tabular}{l|l|rr|rr|rr|rr|rr}
\xl{2}
\multirow{2}{*}{Permission} & \multirow{2}{*}{Type} & \multicolumn{2}{c|}{{2019}} & \multicolumn{2}{c|}{{2020}} & \multicolumn{2}{c|}{{2021}} & \multicolumn{2}{c|}{{2022}} & \multicolumn{2}{c}{{2023}} \\ 
\cline{3-12}
& & {B}  & {M}   & {B}  & {M}   & {B}  & {M}   & {B}  & {M}   & {B}  & {M}   \\ 
\xl{1}
{\tt ACCESS\_NETWORK\_STATE} & G2 & 551 & 99 & 639 & 59 & 726 & 66 & 892 & 58 & 785 & 28 \\
{\tt ACCESS\_FINE\_LOCATION} & G4 & 533 & 101 & 661 & 59 & 820 & 71 & 1,035 & 60 & 886 & 31 \\
{\tt ACCESS\_COARSE\_LOCATION} & G4 & 520 & 101 & 644 & 59 & 810 & 71 & 1,018 & 60 & 867 & 30 \\
{\tt INTERNET} & G2 & 451 & 87 & 516 & 41 & 560 & 52 & 615 & 48 & 534 & 20 \\
{\tt WRITE\_EXTERNAL\_STORAGE} & G3 & 450 & 75 & 488 & 44 & 538 & 60 & 553 & 51 & 433 & 19 \\
{\tt UPDATE\_DEVICE\_STATS} & G2 & 424 & 84 & 515 & 51 & 664 & 63 & 906 & 55 & 811 & 25 \\
{\tt CAMERA} & G6 & 305 & 49 & 431 & 33 & 496 & 53 & 526 & 0 & 0 & 0\\
{\tt RECORD\_AUDIO} & G6 & 273 & 45 & 386 & 31 & 0 & 51 & 0 & 0 & 0 & 0 \\
{\tt USE\_FINGERPRINT} & G7 & 255 & 46 & 416 & 30 & 0 & 0 & 0 & 0 & 0 & 0 \\
{\tt WAKE\_LOCK} & G1 & 254 & 0 & 396 & 33 & 530 & 51 & 744 & 53 & 695 & 23 \\
{\tt READ\_PHONE\_STATE} & G1 & 0 & 44 & 0 & 0 & 0 & 0 & 0 & 44 & 0 & 0 \\
{\tt STATUS\_BAR\_SERVICE} & G8 & 0 & 0 & 0 & 0 & 511 & 0 & 670 & 0 & 613 & 18 \\
{\tt MEDIA\_CONTENT\_CONTROL} & G6 & 0 & 0 & 0 & 0 & 511 & 0 & 670 & 0 & 614 & 18 \\
{\tt GET\_ACCOUNTS} & G1 & 0 & 0 & 0 & 0 & 0 & 42 & 0 & 46 & 449 & 19 \\
{\tt READ\_EXTERNAL\_STORAGE} & G3 & 0 & 0 & 0 & 0 & 0 & 0 & 0 & 45 & 0 & 0\\ 
\xl{2}
\end{tabular}}\vspace{-3mm}
\end{table}

\begin{takeaway}
Malicious apps requested less sensitive permissions over time, possibly to evade detection. On the other hand, benign apps frequently request sensitive permissions, highlighting ongoing considerations for user privacy and security.
\end{takeaway}

\subsubsection{Permissions by App Genre}
We also examined the top requested permissions across 16 app genres, including \emph{finance},~\emph{business},~\emph{education}, and more. This analysis aimed to understand how permission requests vary by app category, as illustrated in~\autoref{tab:permissions_per_genre1-8} and \autoref{tab:permissions_per_genre8-16}. 

{\em Games} and {\em Other} apps consistently requested the most permissions, especially those falling under ``location and GPS'' (G4), ``network and connectivity'' (G2), and ``system and device management'' (G1). These app categories rely heavily on real-time features such as interactive gameplay and dynamic content access.

In contrast, genres such as {\em Books}, {\em Music}, and {\em Travel} consistently requested fewer permissions, particularly among malicious apps, likely due to their offline functionality and limited access to sensitive data. Across all genres, benign apps requested more permissions than malicious ones. Commonly requested permissions among benign apps included {\tt ACCESS\_FINE\_LOCATION}, {\tt WRITE\_EXTERNAL\_ STORAGE}, and {\tt INTERNET}, which fall under G4, G3, and G2, respectively. On the other hand, malicious apps tended to minimize permission requests, likely as an evasion tactic. Interestingly, permissions under ``communication and messaging'' (G5), such as {\tt SEND\_SMS}, {\tt CALL\_PHONE}, and {\tt READ\_CONTACTS}, appeared exclusively in malicious apps and primarily within the {\em Games} category, suggesting potential misuse for spam or fraud activities.
\begin{takeaway}
Permission patterns vary significantly by app category. Data-heavy and interactive genres like {\em Games} and {\em Communication} request more permissions, particularly among benign apps. Malicious apps tend to stay minimal across the board, though some sensitive permissions appear exclusively within certain categories.
\end{takeaway}

\begin{table*}[t]
\centering
\caption{Breakdown of top permissions requested in the first 8 of 16 total app categories for benign (B) and malicious (M). }\label{tab:permissions_per_genre1-8} 
\vspace{-3mm}
\scalebox{0.85}{
\begin{tabular}{l|l|rr|rr|rr|rr|rr|rr|rr|rr}
\xl{2}
\multirow{2}{*}{Permission} & \multirow{2}{*}{Type}  & \multicolumn{2}{c}{Finance} & \multicolumn{2}{c}{Business} & \multicolumn{2}{c}{Education} & \multicolumn{2}{c}{Tools} & \multicolumn{2}{c}{Productivity} & \multicolumn{2}{c}{Lifestyle} & \multicolumn{2}{c}{Medical} & \multicolumn{2}{c}{Books} \\
\cline{3-18}
 &  & B & M & B & M & B & M & B & M & B & M & B & M & B & M & B & M \\
\xl{1}
{\tt ACCESS\_FINE\_LOCATION}   & G4 & 393 & 26 & 478 & 23 & 358 & 31 & 295 & 17 & 277 & 19 & 164 & 5 & 205 & 11 & 99 & 7 \\
{\tt ACCESS\_COARSE\_LOCATION} & G4 & 389 & 25 & 475 & 23 & 349 & 31 & 290 & 17 & 276 & 19 & 157 & 5 & 196 & 11 & 98 & 7 \\
{\tt ACCESS\_NETWORK\_STATE}   & G2 & 386 & 28 & 366 & 21 & 341 & 30 & 242 & 15 & 210 & 14 & 147 & 5 & 171 & 11 & 101 & 6 \\
{\tt UPDATE\_DEVICE\_STATS}    & G1 & 359 & 27 & 381 & 17 & 295 & 24 & 218 & 12 & 234 & 13 & 136 & 4 & 165 & 9 & 88 & 5 \\
{\tt WAKE\_LOCK}               & G1 & 322 & 24 & 286 & 18 & 234 & 0 & 163 & 8 & 169 & 0 & 118 & 5 & 125 & 6 & 75 & 3 \\
{\tt INTERNET}                 & G2 & 297 & 23 & 163 & 13 & 233 & 23 & 177 & 12 & 114 & 11 & 108 & 0 & 120 & 8 & 94 & 4 \\
{\tt CAMERA}                   & G6 & 254 & 0 & 134 & 12 & 193 & 21 & 168 & 10 & 0 & 10 & 78 & 0 & 95 & 7 & 78 & 0 \\
{\tt WRITE\_EXTERNAL\_STORAGE} & G3 & 232 & 15 & 0 & 13 & 218 & 27 & 198 & 14 & 92 & 10 & 91 & 0 & 84 & 10 & 93 & 3 \\
{\tt READ\_PHONE\_STATE}       & G1 & 216 & 0 & 0 & 0 & 0 & 18 & 0 & 0 & 0 & 0 & 0 & 4 & 0 & 9 & 0 & 0 \\
{\tt MEDIA\_CONTENT\_CONTROL}  & G6 & 214 & 18 & 361 & 11 & 217 & 0 & 0 & 0 & 213 & 10 & 97 & 5 & 135 & 0 & 0 & 0 \\
{\tt GET\_ACCOUNTS}            & G3 & 0 & 19 & 0 & 0 & 0 & 0 & 0 & 8 & 0 & 0 & 0 & 4 & 0 & 0 & 0 & 0 \\
{\tt STATUS\_BAR\_SERVICE}     & G8 & 0 & 18 & 361 & 0 & 217 & 0 & 150 & 0 & 213 & 10 & 97 & 5 & 135 & 0 & 0 & 3 \\
{\tt USE\_FINGERPRINT}         & G7 & 0 & 0 & 190 & 11 & 0 & 0 & 0 & 0 & 87 & 0 & 0 & 0 & 0 & 0 & 0 & 4 \\
{\tt RECORD\_AUDIO}            & G6 & 0 & 0 & 0 & 0 & 0 & 19 & 0 & 10 & 0 & 0 & 0 & 0 & 0 & 6 & 72 & 0 \\
{\tt READ\_EXTERNAL\_STORAGE}  & G3 & 0 & 0 & 0 & 0 & 0 & 19 & 152 & 0 & 0 & 11 & 0 & 4 & 0 & 0 & 62 & 0 \\
{\tt RECEIVE\_BOOT\_COMPLETED} & G1 & 0 & 0 & 0 & 0 & 0 & 0 & 0 & 0 & 0 & 0 & 0 & 0 & 0 & 0 & 0 & 3 \\                  
\xl{2}
\end{tabular}}
\end{table*}

\begin{table*}[t]
\centering
\caption{Breakdown of permission requests in benign and malicious android applications across the second set of 8 app categories, completing the overview of all 16 genres from 2019 to 2023 for benign (B) and malicious (M).}\label{tab:permissions_per_genre8-16} 
\vspace{-3mm}
\scalebox{0.95}{
\begin{tabular}{l|l|rr|rr|rr|rr|rr|rr|rr|rr}
\xl{2}
\multirow{2}{*}{Permission} & \multirow{2}{*}{Type}  & \multicolumn{2}{c}{Shopping} & \multicolumn{2}{c}{Enter.} & \multicolumn{2}{c}{Sports} & \multicolumn{2}{c}{Music} & \multicolumn{2}{c}{Travel} & \multicolumn{2}{c}{Comm} & \multicolumn{2}{c}{Games} & \multicolumn{2}{c}{Other} \\
\cline{3-18}
 &  & B & M & B & M & B & M & B & M & B & M & B & M & B & M & B & M \\
\xl{1}
{\tt ACCESS\_FINE\_LOCATION} & G4 & 123 & 15 & 112 & 10 & 117 & 6 & 107 & 14 & 70 & 6 & 88 & 7 & 533 & 63 & 490 & 56 \\
{\tt ACCESS\_COARSE\_LOCATION} & G4 & 123 & 15 & 112 & 10 & 117 & 6 & 106 & 14 & 61 & 6 & 88 & 7 & 515 & 63 & 485 & 56 \\
{\tt ACCESS\_NETWORK\_STATE} & G2 & 115 & 15 & 102 & 10 & 107 & 6 & 107 & 14 & 63 & 6 & 86 & 8 & 546 & 62 & 481 & 54 \\
{\tt UPDATE\_DEVICE\_STATS} & G1 & 114 & 15 & 89 & 10 & 104 & 5 & 92 & 12 & 51 & 5 & 74 & 0 & 505 & 63 & 393 & 47 \\
{\tt WAKE\_LOCK} & G1 & 101 & 12 & 68 & 0 & 90 & 0 & 0 & 0 & 42 & 3 & 69 & 0 & 365 & 40 & 334 & 34 \\
{\tt INTERNET} & G2 & 97 & 9 & 94 & 9 & 78 & 5 & 98 & 11 & 45 & 4 & 0 & 0 & 504 & 61 & 385 & 42 \\
{\tt CAMERA} & G6 & 71 & 8 & 79 & 9 & 58 & 5 & 82 & 9 & 42 & 3 & 48 & 7 & 404 & 38 & 293 & 29 \\
{\tt WRITE\_EXTERNAL\_STORAGE} & G3 & 74 & 0 & 91 & 9 & 72 & 6 & 102 & 14 & 43 & 4 & 52 & 7 & 525 & 59 & 359 & 43 \\
{\tt READ\_PHONE\_STATE} & G1 & 0 & 0 & 69 & 6 & 0 & 0 & 68 & 0 & 0 & 0 & 0 & 7 & 0 & 32 & 0 & 0 \\
{\tt MEDIA\_CONTENT\_CONTROL} & G6 & 66 & 0 & 0 & 0 & 78 & 0 & 0 & 0 & 38 & 0 & 56 & 0 & 0 & 0 & 0 & 0 \\
{\tt GET\_ACCOUNTS} & G3 & 0 & 7 & 0 & 0 & 0 & 0 & 0 & 0 & 0 & 3 & 0 & 0 & 0 & 35 & 249 & 27 \\
{\tt STATUS\_BAR\_SERVICE} & G8 & 65 & 0 & 0 & 0 & 78 & 0 & 0 & 0 & 38 & 0 & 56 & 0 & 0 & 0 & 0 & 0 \\
{\tt USE\_FINGERPRINT} & G7 & 0 & 0 & 0 & 0 & 0 & 0 & 0 & 7 & 0 & 0 & 0 & 0 & 0 & 0 & 0 & 0 \\
{\tt RECORD\_AUDIO} & G6 & 0 & 0 & 0 & 9 & 0 & 5 & 86 & 9 & 0 & 0 & 0 & 7 & 393 & 0 & 0 & 0 \\
{\tt READ\_EXTERNAL\_STORAGE} & G3 & 0 & 0 & 74 & 8 & 0 & 4 & 78 & 9 & 0 & 3 & 49 & 6 & 0 & 0 & 287 & 35 \\
{\tt RECEIVE\_BOOT\_COMPLETED} & G1 & 0 & 0 & 0 & 0 & 0 & 0 & 0 & 0 & 0 & 0 & 0 & 0 & 0 & 0 & 0 & 0 \\
{\tt ACCESS\_WIFI\_STATE} & G2 & 0 & 8 & 0 & 0 & 0 & 0 & 0 & 0 & 0 & 0 & 0 & 0 & 362 & 0 & 0 & 0 \\
{\tt CALL\_PHONE} & G5 & 0 & 7 & 0 & 0 & 0 & 0 & 0 & 0 & 0 & 0 & 0 & 0 & 0 & 0 & 0 & 0 \\
{\tt BLUETOOTH} & G5 & 0 & 0 & 0 & 0 & 0 & 3 & 0 & 0 & 0 & 0 & 0 & 0 & 0 & 0 & 0 & 0 \\
{\tt SEND\_SMS} & G5 & 0 & 0 & 0 & 0 & 0 & 0 & 0 & 0 & 0 & 0 & 0 & 7 & 0 & 0 & 0 & 0 \\
{\tt READ\_CONTACTS} & G5 & 0 & 0 & 0 & 0 & 0 & 0 & 0 & 0 & 0 & 0 & 0 & 7 & 0 & 0 & 0 & 0 \\    
\xl{2}
\end{tabular}}
\end{table*}

\subsection{Association Rule Analysis}
Association rule analysis is a powerful data mining technique used to uncover relationships between items in large datasets. A core component of this technique is Frequent Itemset Mining (FIM), which focuses on discovering sets of items that frequently appear together. In this work, we applied FIM to identify the most common combinations of Android permissions requested by apps. Our primary objective was to uncover these co-occurrence patterns to understand how app behaviors differ across categories, years, and between malicious and benign applications, rather than creating predictive rules about one permission's presence implying another.

Our analysis centers on the support metric, which directly measures the frequency of these permission sets. While other metrics like confidence and lift are essential for evaluating the predictive strength of association rules, they were omitted because our goal was to describe which permissions co-occur frequently, not to predict the presence of one permission from another. This focus on frequent itemsets allows us to directly identify the characteristic permission bundles that define different classes of applications, providing a deeper understanding of permission usage dynamics and their implications for user privacy and security.

To achieve this, we employed the \textit{FP-Growth (Frequent Pattern Growth)} algorithm, which is a highly efficient method for finding frequent itemsets in large datasets. Unlike other algorithms like Apriori, FP-Growth does not generate candidate itemsets explicitly. Instead, it builds a compressed data structure called the Frequent Pattern Tree (FP-Tree), which stores information about item frequencies in a hierarchical format. The algorithm then mines the FP-Tree to discover frequent itemsets. The formula used to determine the frequency of an itemset is defined as:

\begin{equation}
\text{Support} = \frac{\text{Frequency of the Itemset in Dataset}}{\text{Total Number of Transactions in Dataset}}.
\label{eq:support}
\end{equation}


In this study, we applied a minimum support threshold of 50\% for the FP-Growth algorithm, meaning that a permission combination was included only if it appeared in at least half of the analyzed applications. This threshold was chosen to highlight the most dominant and widely shared permission patterns across apps, ensuring meaningful insights while filtering out low-frequency or less relevant combinations. The selection of a 50\% threshold aligns with common practice in association rule mining, particularly in studies aiming to extract high-confidence and interpretable patterns \cite{HanKP2011, AgrawalIS93}. By focusing on frequent associations, we aim to surface stable trends rather than outliers. A detailed example is provided to illustrate how the FP-Growth algorithm operates in this context.

\begin{figure}[t]
    \centering
    \includegraphics[width=0.99\linewidth]{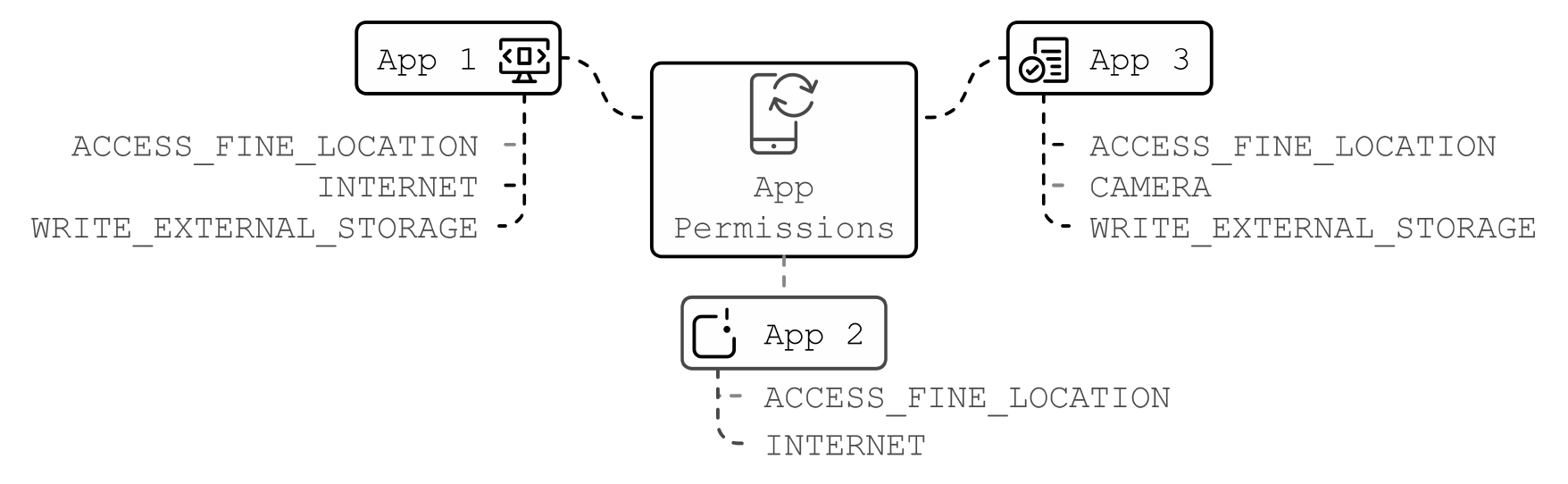}
    \caption{Example of permission sets for three apps used to demonstrate the FP-Growth algorithm.}
    \label{fig:permissions-fpgrowth}\vspace{-3mm}
\end{figure}

The first step of the FP-Growth algorithm is to calculate the support for individual permissions. For instance, the permission \texttt{ACCESS\_FINE\_LOCATION} appears in all three apps, resulting in a support value of \(3/3 = 100\%\). Similarly, \texttt{INTERNET} appears in two apps, yielding a support of \(2/3 \approx 66.7\%\). Based on the support threshold of 50\%, all these permissions would be included in the FP-Tree. Next, the algorithm identifies frequent permission combinations. For example, the combination {\tt {ACCESS\_FINE\_LOCATION, INTERNET}} appears in two apps, resulting in a support of \(2/3 \approx 66.7\%\). This process continues until all frequent combinations of permissions are identified. For our analysis, we applied the FP-Growth algorithm to three distinct dimensions of the dataset. First, we examined the entire dataset to identify frequent permission combinations across all malicious and benign apps, providing a comprehensive view of co-occurring permissions. Second, we conducted a yearly analysis from 2019 to 2023, analyzing trends in permission combinations over time by separating malicious and benign apps. Finally, we performed a genre-specific analysis, focusing on 16 distinct app genres, such as Games, Finance, and Communication, to uncover permission patterns unique to each genre.

\subsubsection{Whole dataset Permission Combinations}

To gain insights into permission request patterns, we analyzed the entire dataset, separating malicious and benign apps to identify differences in permission usage. Table \ref{table:fpgrowth_benign_malicious_comparison} presents the findings, organized by \textit{``Permission Size"} refers to the number of permissions grouped together in a combination analyzed through the FP-Growth algorithm. For example, a permission size of 2 could include combinations like {\sloppy\texttt{ACCESS\_\allowbreak FINE\_LOCATION}}, {\tt INTERNET}, where two permissions co-occur in the dataset. Similarly, a permission size of 3 might include 
{\sloppy\texttt{ACCESS\_FINE\_\allowbreak LOCATION}}, {\tt INTERNET}, {\tt WRITE\_EXTERNAL\_STORAGE}, indicating three permissions commonly requested together. The \textit{``Avg. Support (\%)"} column indicates how often these permission combinations occur across the respective app types, calculated as the percentage of apps in the dataset containing these combinations. The \textit{``Total Count"} column quantifies the number of apps that include these permission combinations, providing a tangible measure of their prevalence.

Benign apps tend to exhibit larger and more diverse permission combinations, as reflected by both the higher average support percentages and the higher total counts for combinations of larger permission sizes. For example, two-permission combinations in benign apps have an average support of 61.41\%, with a total count of 58,136. This total count represents the sum of occurrences of all unique two-permission combinations across all benign apps in the dataset, not the number of apps themselves. In contrast, malicious apps for the same permission size have an average support of 56.61\% and a total count of 895, indicating significantly fewer instances of two-permission combinations. As the permission size increases, this gap becomes more apparent, with benign apps maintaining high counts and supports for larger combinations, while malicious apps show fewer or no frequent itemsets for sizes beyond three permissions. This suggests that malicious apps often employ a more targeted and minimalistic approach to requesting permissions, potentially to avoid detection.

\begin{table}[t]
\renewcommand{\arraystretch}{0.9}
\centering
\caption{Comparison of benign and malicious applications: average support and total count by permission size. Permission size refers to the number of permissions grouped together, with Average Support showing the average percentage of apps requesting each combination and total count representing the overall frequency of these combinations.}
\label{table:fpgrowth_benign_malicious_comparison}\vspace{-2mm}
\begin{tabular}{c|r|r|r|r}
\hline
\multirow{2}{*}{Permission Size} & \multicolumn{2}{c|}{Avg. Support (\%)} & \multicolumn{2}{c}{Total Count} \\ \cline{2-5} 
                      & Benign & Malicious & Benign & Malicious \\ \hline
2                     & 61.41  & 56.61     & 58,136 & 895       \\ 
3                     & 57.90  & 55.22     & 57,304 & 291       \\ 
4                     & 55.72  & -         & 26,374 & -         \\ 
5                     & 53.94  & -         & 4,642  & -         \\ 
\cline{1-5} 
\end{tabular}
\end{table}

\begin{takeaway}
Benign apps request broader permissions, while malicious apps focus on minimal combinations, likely to evade detection.
\end{takeaway}

\subsubsection{Yearly Permission Combinations}

The yearly analysis highlights the evolution of permission combinations from 2019 to 2023 across benign and malicious applications. Table \ref{table:fpgrowth_yearly_benign_malicious_comparison} presents the average support percentages and total counts of permission combinations for different permission sizes over these years. The data shows that benign apps consistently request a greater number of larger permission combinations compared to malicious apps. For instance, in 2019, benign apps exhibited an average support of 67.51\% for two-permission combinations, encompassing 6,471 instances, compared to 66.61\% and 1,229 instances for malicious apps. However, as the permission size increases, the support percentages and total counts for both benign and malicious apps decline. By 2022, benign apps maintained 7,078 four-permission combinations with an average support of 58.87\%, while malicious apps recorded only 50 such combinations with a support of 52.08\%.

A notable trend observed is the diminishing frequency of larger permission combinations in malicious apps over time. For example, while benign apps frequently recorded combinations of four and five permissions, malicious apps increasingly focused on smaller permission sets beyond 2021. This shift could suggest a strategic move by malicious apps to minimize detection by avoiding excessive requests. The yearly breakdown also reveals the stability of benign app behavior over time, with consistent patterns in permission requests, particularly for larger combinations. Malicious apps, on the other hand, exhibit a more pronounced decline in support and total counts for larger combinations as the years progress. This distinction highlights evolving strategies in permission requests, with malicious apps adopting a more streamlined approach.

\begin{table}[t]
\renewcommand{\arraystretch}{0.9}
\centering
\caption{Comparison of benign (B) and malicious (M) apps by year: average support and total count by permission size.}
\label{table:fpgrowth_yearly_benign_malicious_comparison}\vspace{-3mm}
\scalebox{0.90}{
\begin{tabular}{c|c|rr|rr}
\hline
\multirow{2}{*}{Year} & \multirow{2}{*}{Permission Size} & \multicolumn{2}{c|}{Avg. Support (\%)} & \multicolumn{2}{c}{Total Count} \\ \cline{3-6}
                      &                                & B & M & B & M \\ \hline
\multirow{5}{*}{2019}  & 2                              & 67.51  & 66.61     & 6,471  & 1,229     \\ 
                       & 3                              & 61.82  & 61.22     & 7,901  & 1,506     \\ 
                       & 4                              & 57.83  & 57.29     & 5,543  & 1,057     \\ 
                       & 5                              & 54.85  & 54.34     & 2,103  & 401       \\ 
                       & 6                              & 52.43  & 52.03     & 335    & 64        \\ \hline
\multirow{4}{*}{2020}  & 2                              & 60.72  & 57.33     & 14,164 & 313       \\ 
                       & 3                              & 56.63  & 54.12     & 18,578 & 197       \\ 
                       & 4                              & 54.53  & 51.65     & 11,130 & 47        \\ \hline
\multirow{4}{*}{2021}  & 2                              & 60.80  & 54.62     & 14,274 & 65        \\ 
                       & 3                              & 57.74  & -         & 14,078 & -         \\ 
                       & 4                              & 55.91  & -         & 6,058  & -         \\ 
                       & 5                              & 53.82  & -         & 972    & -         \\ \hline
\multirow{3}{*}{2022}  & 2                              & 65.26  & 53.31     & 14,266 & 563       \\ 
                       & 3                              & 61.17  & 52.60     & 14,710 & 303       \\ 
                       & 4                              & 58.87  & 52.08     & 7,078  & 50        \\ \hline
\multirow{5}{*}{2023}  & 2                              & 66.54  & -         & 14,370 & -         \\ 
                       & 3                              & 61.20  & -         & 17,814 & -         \\ 
                       & 4                              & 57.96  & -         & 10,884 & -         \\ 
                       & 5                              & 54.30  & -         & 3,569  & -         \\ 
                       & 6                              & 50.59  & -         & 475    & -         \\ \hline
\end{tabular}}
\end{table}

\begin{takeaway}
Benign apps continue to use larger permission combinations, while malicious apps increasingly focus on smaller sets, reflecting a strategic shift to minimize detection.
\end{takeaway}

\subsubsection{Genres Permission Combinations}

Our genre-specific analysis examined frequent permission combinations across 16 app genres, distinguishing between benign and malicious applications. The results, presented in Tables \ref{table:1-8genres_comparison} and \ref{table:8-16genres_comparison}, provide insights into how permission requests vary based on app categories, offering a nuanced perspective on app behavior.

Benign apps consistently showed higher total counts across all permission sizes in most genre, reflecting broader functionality and more diverse access requirements. For example, in the {\em Games} genre, benign apps had 18,653 permission sets of size 2 with an average support of 68.15\%, while malicious apps had just 1,007. Despite the lower count, the malicious apps showed a slightly higher average support of 72.66\%, which suggests more targeted combinations.

Genres like {\em Books}, {\em Entertainment}, {\em Business}, and {\em Travel} had benign apps requesting more complex permissions. In contrast, malicious apps were more concentrated in high-risk categories like {\em Communication}, {\em Games}, {\em Music}, {\em Shopping}, and {\em Sports}. These categories involve frequent user interactions, access to personal data and financial features, which make them appealing targets for attackers.

The {\em Communication} genre stood out, with some malicious apps requesting permission sets as large as size 11. This indicates the potential for highly invasive behaviors. Meanwhile, genres like {\em Medical} and {\em Tools} showed little to no significant malicious activity, pointing to generally lower risk levels. This analysis suggests that attackers may strategically choose certain genres where users are more likely to grant permissions without suspicion. Developers and users alike must be aware of these trends to better manage risk.

\begin{table}[t]
\renewcommand{\arraystretch}{0.9}
\centering
\caption{Comparison of average support and total count by permission size for benign (B) and malicious (M) apps across the first 8 of 16 total app genres.}
\label{table:1-8genres_comparison}\vspace{-3mm}
\scalebox{0.90}{
\begin{tabular}{c|c|cc|cc}
\hline
\multirow{2}{*}{\text{Genre}} & \multirow{2}{*}{\text{Permission Size}} & \multicolumn{2}{c|}{Avg. Support (\%)} & \multicolumn{2}{c}{Total Count} \\ \cline{3-6}
 & & B & M & B & M \\ \hline

\multirow{9}{*}{Books} 
& 2 & 66.02 & 55.56 & 2,699 & 15 \\
& 3 & 62.54 & 55.56 & 6,255 & 5 \\
& 4 & 59.97 & - & 9,544 & - \\
& 5 & 57.98 & - & 9,778 & - \\
& 6 & 56.41 & - & 6,630 & - \\
& 7 & 55.08 & - & 2,855 & - \\
& 8 & 53.88 & - & 708 & - \\
& 9 & 52.74 & - & 77 & - \\ \hline

\multirow{6}{*}{Business} 
& 2 & 67.97 & 58.47 & 6,341 & 145 \\
& 3 & 61.57 & 56.99 & 7,860 & 53 \\
& 4 & 57.48 & - & 5,362 & - \\
& 5 & 54.76 & - & 1,882 & - \\
& 6 & 52.95 & - & 260 & - \\ \hline

\multirow{11}{*}{Communication} 
& 2 & 64.22 & 70.94 & 1,208 & 415 \\
& 3 & 59.01 & 68.25 & 1,402 & 1,290 \\
& 4 & 55.99 & 65.93 & 776 & 2,670 \\
& 5 & 54.21 & 63.89 & 161 & 3,864 \\
& 6 & - & 62.09 & - & 3,990 \\
& 7 & - & 60.49 & - & 2,940 \\
& 8 & - & 59.06 & - & 1,515 \\
& 9 & - & 57.78 & - & 520 \\
& 10 & - & 56.61 & - & 107 \\
& 11 & - & 55.56 & - & 10 \\ \hline

\multirow{5}{*}{Education} 
& 2 & 61.41 & - & 6,334 & - \\
& 3 & 57.41 & - & 6,798 & - \\
& 4 & 54.92 & - & 3,357 & - \\
& 5 & 53.14 & - & 609 & - \\ \hline

\multirow{7}{*}{Entertainment} 
& 2 & 61.65 & - & 3,535 & - \\
& 3 & 57.86 & - & 6,424 & - \\
& 4 & 55.46 & - & 6,699 & - \\
& 5 & 53.64 & - & 4,057 & - \\
& 6 & 52.11 & - & 1,335 & - \\
& 7 & 50.82 & - & 186 & - \\ \hline

\multirow{4}{*}{Finance} 
& 2 & 63.08 & 55.27 & 5,439 & 325 \\
& 3 & 59.44 & 52.78 & 5,694 & 266 \\
& 4 & 56.30 & 51.59 & 3,236 & 65 \\ 
& 5 & 54.07 & - & 777 & - \\ \hline

\multirow{8}{*}{Games} 
& 2 & 68.15 & 72.66 & 18,653 & 1,007 \\
& 3 & 63.50 & 72.85 & 38,159 & 1,234 \\
& 4 & 60.48 & 73.68 & 46,779 & 851 \\
& 5 & 58.35 & 72.51 & 35,064 & 335 \\
& 6 & 56.76 & 71.43 & 15,536 & 55 \\
& 7 & 55.54 & - & 3,635 & - \\
& 8 & 54.62 & - & 325 & - \\ \hline

\multirow{5}{*}{Lifestyle} 
& 2 & 64.60 & - & 2,274 & - \\
& 3 & 61.90 & - & 2,179 & - \\
& 4 & 59.66 & - & 1,050 & - \\
& 5 & 57.67 & - & 203 & - \\ \hline

\end{tabular}
}\vspace{-3mm}
\end{table}

\begin{table}[t]\renewcommand{\arraystretch}{0.9}
\centering
\caption{Comparison of average support and total count by permission size for benign (B) and malicious (M) apps across the second 8 of 16 total app genres.}
\label{table:8-16genres_comparison}\vspace{-3mm}
\scalebox{0.90}{
\begin{tabular}{c|c|cc|cc}
\hline
\multirow{2}{*}{\text{Genre}} & \multirow{2}{*}{\text{Permission Size}} & \multicolumn{2}{c|}{Avg. Support (\%)} & \multicolumn{2}{c}{Total Count} \\ \cline{3-6}
 & & B & M & B & M \\ \hline

\multirow{5}{*}{Medical} 
& 2 & 60.87 & - & 2,715 & - \\
& 3 & 56.22 & - & 2,758 & - \\
& 4 & 53.40 & - & 1,310 & - \\
& 5 & 51.35 & - & 229 & - \\ \hline

\multirow{9}{*}{Music} 
& 2 & 63.32 & 62.94 & 5,243 & 321 \\
& 3 & 60.37 & 59.17 & 11,456 & 513 \\
& 4 & 58.09 & 56.66 & 16,166 & 472 \\
& 5 & 56.29 & 54.90 & 15,018 & 252 \\
& 6 & 54.92 & 53.68 & 8,968 & 73 \\
& 7 & 53.85 & 52.94 & 3,282 & 9 \\
& 8 & 52.96 & - & 670 & - \\
& 9 & 52.17 & - & 60 & - \\ \hline

\multirow{5}{*}{Other} 
& 2 & 61.70 & 56.13 & 7,975 & 293 \\
& 3 & 58.23 & 52.87 & 8,182 & 184 \\
& 4 & 55.80 & 50.57 & 4,077 & 44 \\
& 5 & 53.38 & - & 900 & - \\ \hline

\multirow{5}{*}{Productivity} 
& 2 & 68.09 & 63.48 & 3,726 & 73 \\
& 3 & 62.07 & 58.70 & 4,290 & 27 \\
& 4 & 58.61 & - & 2,363 & - \\
& 5 & 56.60 & - & 489 & - \\ \hline

\multirow{9}{*}{Shopping} 
& 2 & 68.04 & 65.50 & 3,952 & 262 \\
& 3 & 62.07 & 58.84 & 8,848 & 386 \\
& 4 & 58.42 & 54.33 & 12,493 & 339 \\
& 5 & 56.04 & 51.42 & 11,466 & 181 \\
& 6 & 54.37 & 50.00 & 6,890 & 56 \\
& 7 & 53.13 & 50.00 & 2,665 & 8 \\
& 8 & 52.19 & - & 620 & - \\
& 9 & 51.52 & - & 68 & - \\ \hline

\multirow{8}{*}{Sports} 
& 2 & 65.49 & 68.98 & 2,497 & 169 \\
& 3 & 59.64 & 64.00 & 3,888 & 336 \\
& 4 & 55.93 & 60.71 & 3,440 & 408 \\
& 5 & 53.30 & 58.67 & 1,770 & 308 \\
& 6 & 51.42 & 57.55 & 506 & 141 \\
& 7 & 50.41 & 57.14 & 62 & 36 \\
& 8 & - & 57.14 & - & 4 \\ \hline

\multirow{4}{*}{Tools} 
& 2 & 59.51 & - & 2,471 & - \\
& 3 & 54.62 & - & 1,890 & - \\
& 4 & 51.64 & - & 536 & - \\ \hline

\multirow{4}{*}{Travel} 
& 2 & 60.51 & - & 1,350 & - \\
& 3 & 57.51 & - & 1,283 & - \\
& 4 & 55.67 & - & 594 & - \\
& 5 & 54.12 & - & 105 & - \\ \hline

\end{tabular}
}
\end{table}

\begin{takeaway}
Benign apps tend to request larger permission sets in content-rich categories, while malicious apps concentrate in high-risk genres where they can exploit user trust and access sensitive data.
\end{takeaway}

\BfPara{Interpreting Frequent Permission Patterns} To further contextualize our findings, we examined a few frequent permission combinations and explored their potential functional justifications. For instance, the combination \texttt{ACCESS\_FINE\_ LOCATION} and \texttt{INTERNET} was among the most common in benign apps. This pairing is functionally justified in apps like navigation, delivery, or weather applications, which require precise location data and internet connectivity to provide real-time updates or services. Another frequent trio observed in benign apps, \texttt{READ\_EXTERNAL\_ STORAGE}, \texttt{WRITE\_EXTERNAL\_STORAGE}, and \texttt{INTERNET}, is typical in media or file-sharing applications, where users download and store content on the device. In contrast, some permission combinations in malicious apps suggest potentially suspicious or privacy-invasive behavior. For example, the combination \texttt{READ\_SMS} and \texttt{READ\_CONTACTS} is frequently seen in messaging trojans or phishing malware, where the goal is to harvest sensitive communication data and user contact lists. When paired with \texttt{INTERNET}, this enables data exfiltration to external servers. These examples illustrate how frequent itemsets can reflect both legitimate app functionalities and potential vectors for misuse, reinforcing the value of permission combination analysis for security diagnostics.

\subsection{Comparative Analysis}

\subsubsection{Permissions with Ads}
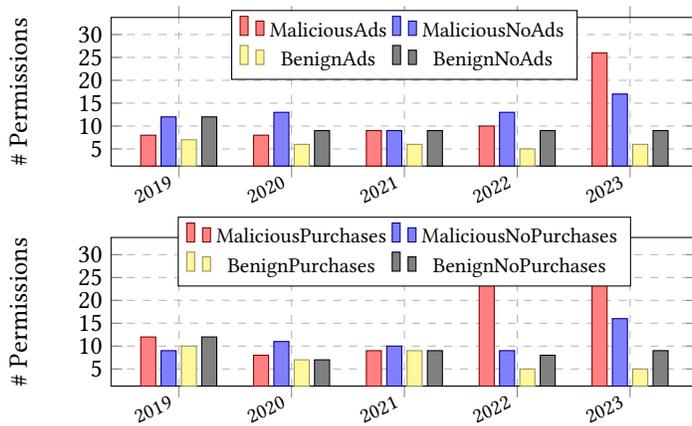
\begin{figure}[t]
\pgfplotsset{}
   \centering
\begin{tikzpicture}
    \begin{axis}
    [
        ybar,
        ymin = 5, ymax = 30,
        ytick distance = 5,
        enlargelimits=0.15,
        width = 0.5\textwidth,
        height = 0.20\textwidth,
        ylabel={\# Permissions},
        ylabel style ={font = \fontsize{10}{12}\selectfont},
        yticklabel style={font = \fontsize{10}{12}\selectfont},
        symbolic x coords={2019, 2020, 2021, 2022, 2023},
        x=1.5cm,
        bar width=2mm,
        legend style={at={(0.5,1.05)},anchor=north, legend columns=2, font=\fontsize{8}{10}\selectfont},
        xticklabel style={rotate=25,anchor=east, align=right, font=\fontsize{8}{10}\selectfont},
        grid=both,grid style={line width=0.1pt, draw=gray!10},major grid style={line width=.1pt,draw=black!30, dashed}
        ]       
    \addplot[ybar, fill=red!50!white, draw=red!50!black] coordinates {(2019,8) (2020,8) (2021,9) (2022,10) (2023,26)};
     \addlegendentry{MaliciousAds};
    \addplot[ybar, fill=blue!50!white, draw=blue!50!black] coordinates {(2019,12) (2020,13) (2021,9) (2022,13) (2023,17)};
    \addlegendentry{MaliciousNoAds};
    \addplot[ybar, fill=yellow!50!white, draw=yellow!50!black] coordinates {(2019,7) (2020,6) (2021,6) (2022,5) (2023,6)};
     \addlegendentry{BenignAds};
    \addplot[ybar, fill=black!50!white, draw=black!50!black] coordinates {(2019,12) (2020,9) (2021,9) (2022,9) (2023,9)};
     \addlegendentry{BenignNoAds};
    \end{axis}
\end{tikzpicture}

\pgfplotsset{}
   \centering
\begin{tikzpicture}
    \begin{axis}
    [
        ybar,
        ymin = 5, ymax = 30,
        ytick distance = 5,
        enlargelimits=0.15,
        width = 0.5\textwidth,
        height = 0.20\textwidth,
        ylabel={\# Permissions},
        ylabel style ={font = \fontsize{10}{12}\selectfont},
        yticklabel style={font = \fontsize{10}{12}\selectfont},
        symbolic x coords={2019, 2020, 2021, 2022, 2023},
        x=1.5cm,
        bar width=2mm,
        legend style={at={(0.5,1.14)},anchor=north, legend columns=2, font=\fontsize{8}{10}\selectfont},
        xticklabel style={rotate=25,anchor=east, align=right, font=\fontsize{8}{10}\selectfont},
        grid=both,grid style={line width=0.1pt, draw=gray!10},major grid style={line width=.1pt,draw=black!30, dashed}
        ]       
    \addplot[ybar, fill=red!50!white, draw=red!50!black] coordinates {(2019,12) (2020,8) (2021,9) (2022,28) (2023,30)};
     \addlegendentry{MaliciousPurchases};
    \addplot[ybar, fill=blue!50!white, draw=blue!50!black] coordinates {(2019,9) (2020,11) (2021,10) (2022,9) (2023,16)};
    \addlegendentry{MaliciousNoPurchases};
    \addplot[ybar, fill=yellow!50!white, draw=yellow!50!black] coordinates {(2019,10) (2020,7) (2021,9) (2022,5) (2023,5)};
     \addlegendentry{BenignPurchases};
    \addplot[ybar, fill=black!50!white, draw=black!50!black] coordinates {(2019,12) (2020,7) (2021,9) (2022,8) (2023,9)};
     \addlegendentry{BenignNoPurchases};
    \end{axis}
\end{tikzpicture}\vspace{-3mm}
\caption{Number of requested permissions in malicious and benign Android applications: (top) offering in-app purchases and (bottom) with ads support.}
\label{fig:MaliciousandBenignAds}
\label{fig:MaliciousandBenignPurchases}\vspace{-3mm}
\end{figure}

Ads in mobile apps are used to generate revenue by displaying promotional content, often delivered through third-party advertising networks. We analyzed whether apps that support ads request more permissions than those that do not for benign and malicious apps. The results, depicted in~\autoref{fig:MaliciousandBenignAds}, highlight several notable trends and distinctions.

For most years from 2019 to 2023, the difference in permission requests between malicious apps with ads and those without was minimal, as both types consistently requested more permissions than their benign counterparts. However, in 2023, we observed a notable shift: malicious apps with ads began requesting significantly more permissions than those without ads. This sudden jump suggests a potential change in strategy, where ad-supported malicious apps may be leveraging advertising frameworks to justify or obscure excessive permission requests. In contrast, benign apps with ads tended to request fewer permissions than benign apps without ads. This could reflect more careful permission management by developers who monetize through advertising, aiming to maintain user trust and comply with platform guidelines.

Another observation is that even without ads, malicious apps still requested more permissions than benign apps. This indicates that the presence of ads alone doesn't account for permission bloat; rather, the underlying intent of the app plays a larger role. Ads may amplify the issue, especially in malicious apps, but they are not the root cause. These findings align with growing concerns around privacy in mobile systems. Ad-supported apps, particularly in recent years, may exploit permission requests to harvest user data under the guise of advertising functionality. The disparity between benign and malicious ad-supported apps underscores the importance of stringent privacy policies and robust security measures to prevent misuse of permissions~\cite{ZhouWWLZC021,HolavanalliMNRSKZ13}. This highlights the need for stronger regulatory measures, clearer app labeling, and user education on the risks of granting unnecessary permissions, especially in apps that appear benign but include aggressive ad frameworks.

\begin{takeaway}
Across the years, the difference between permissions requested by malicious apps with and without ads is minimal, except in 2023, where apps with ads show a sharp increase in the number of permissions requested.
\end{takeaway}

\subsubsection{In-App Purchases} Refers to transactions made within an app, allowing users to buy additional content, features, or services, such as virtual goods, subscriptions, or premium upgrades. These purchases provide a monetization strategy for developers while enhancing user experience. We analyzed whether apps offering in-app purchases request more permissions than those that do not, for both benign and malicious apps. The results, depicted in~\autoref{fig:MaliciousandBenignPurchases}, highlight several notable trends. 

The results reveal a clear trend: malicious apps with in-app purchases consistently request more permissions than malicious apps without them, with the gap becoming especially pronounced in 2022 and 2023. This spike suggests that malicious developers may be using in-app purchase features as a cover to justify excessive permission access, potentially to harvest sensitive user data or enable hidden behaviors. Benign apps also tended to request more permissions when in-app purchases were present, although the pattern was less consistent and more modest in scale. This is likely due to legitimate functionality needs, such as enabling payment processing, account management, or unlocking premium features. Still, the trend indicates that even benign apps need to manage permission requests carefully to maintain user trust.

It is also notable that, even without in-app purchases, malicious apps generally requested more permissions than benign apps. This suggests that in-app purchases are not the only factor driving permission requests; rather, malicious apps are inherently more aggressive in their access demands. These findings highlight an important privacy concern: the blending of functional and potentially harmful permission usage. While in-app purchases often justify added permissions, malicious apps may exploit this as a disguise for intrusive behavior. This underscores the need for transparency from developers and stricter oversight from app marketplaces and regulators. Users should be cautious of apps requesting broad permissions, especially when paired with in-app monetization features.

\begin{takeaway}
Malicious apps with in-app purchases request more permissions, posing privacy risks, while benign apps ensure transparent justified permission requests to maintain user trust and security.
\end{takeaway}

\subsubsection{Content Ratings} We analyzed permission requests in apps across different content ratings over a five-year interval for both benign and malicious apps, as depicted in~\autoref{tab:age_permission_malicious-benign}, to determine how permission request vary across content rating categories.

Content rating refers to an app's age suitability and content guidelines, providing users with information about the appropriate age group for the app's content. Our analysis categorizes apps into four content rating groups: ``Teen 15-17'', ``Mature 17+'', ``Everyone 10+'', and ``Everyone''. The Google Play Store defines these categories and helps users understand the intended audience of each app. The ``Teen 15-17'' category is for apps suitable for teenagers, ``Mature 17+'' is for adults due to mature content, ``Everyone 10+'' includes apps appropriate for a general audience aged 10 and above, and ``Everyone'' indicates apps suitable for all ages. These ratings help users make decisions about the apps they download and use.

Our analysis revealed that benign apps generally request more permissions than malicious ones across all ratings. This is mainly because there are more benign apps in the dataset, with the highest count appearing in 2023. For malicious apps, 2019 peaked in terms of both app count and permission requests, followed by a steady decline, especially in the ``Teen'' and ``Everyone'' categories. This suggests developers may be scaling back permission requests in these categories to avoid scrutiny.

On the other hand, benign apps showed a sharp increase in both the number of apps and permission requests, especially in the Teen and Everyone categories. This likely reflects the growing complexity of these apps, which may require broader access to device functions to support new features or services.

The ``Everyone'' content ratings contained the highest apps count and the largest overall count of permissions across benign and malicious samples. While this category is intended for general audiences, our findings indicate that many apps within it, particularly benign ones, request a substantial number of permissions. This highlights the need for continued permission oversight as even widely accessible apps may exhibit extensive access to device resources.

These observations reinforce an important principle: developers should limit permission requests to those strictly necessary for core functionality, particularly when targeting children, teens, or broad user groups. In parallel, app marketplaces and regulators should provide and enforce clearer guidelines to ensure that permission usage is both justified and appropriate for the app's intended audience and declared functionality.

\begin{takeaway}
Malicious apps request fewer permissions, while benign apps targeting younger users request more, highlighting the need for better oversight. 

\end{takeaway}

\begin{table*}
\centering
\caption{Requested permissions in benign and malicious apps by content rating categories from 2019 to 2023. Teen (15--17), Mature (M17+), Everyone 10+ (E10+), and Everyone (E). 'A' represents the number of applications for each category, and 'P' indicates the total number of permissions requested by that category.}\label{tab:age_permission_malicious-benign}
\vspace{-3mm}
\scalebox{0.87}{
\begin{tabular}{l|rrrrrrrr|rrrrrrrr}
\xl{2}
\multirow{3}{*}{Year} & \multicolumn{8}{c|}{Benign Apps} & \multicolumn{8}{c}{Malicious Apps} \\
\cline{2-17}
& \multicolumn{2}{c}{15-17} 
& \multicolumn{2}{c}{M17+} 
& \multicolumn{2}{c}{E10+ }
& \multicolumn{2}{c|}{E} 
& \multicolumn{2}{c}{15-17}
& \multicolumn{2}{c}{M17+} 
& \multicolumn{2}{c}{E10+} 
& \multicolumn{2}{c}{E}\\
\cline{2-17}
& A & P & A & P & A & P & A & P & A & P & A & P & A & P & A & P \\
\hline
2019 & 28&285 & 6&92 & 7&71 & 254&1,996 & 25&322 & 3&13 & 3&17 & 94&1,001 \\
2020 & 85&1,284 & 17&240 & 16&218 & 618&8,203 & 9&201 & 5&51 & 0&0 & 73&633 \\
2021 & 76&1,149 & 23&370 & 15&227 & 799&10,284 & 12&112 & 2&5 & 1&33 & 105&967 \\
2022 & 84&1,292 & 22&405 & 29&493 & 973&12,613 & 10&143 & 7&53 & 0&0 & 112&971 \\
2023 & 62&1,045 & 23&362 & 16&271 & 952&11,548 & 3&54 & 5&12 & 2&37 & 91&411 \\
\xl{2}
\end{tabular}
}
\end{table*}


\subsubsection{App Star Ratings} In our analysis, we analyzed permission requests in apps across different star ratings over a five-year interval for both benign and malicious apps as shown in~\autoref{tab:maliciouspermissions_by_rating} and~\autoref{tab:benign_permissions_by_star_ratings}. App star ratings reflect user feedback on the Google Play Store, typically ranging from 1 to 5 stars. They help indicate an app's quality, reliability, and user satisfaction. Some apps may have 0 stars, often because they are newly released or have received little to no user feedback.

Our analysis revealed distinct trends in permission requests across star ratings. In both benign and malicious apps, 0-star apps requested the highest number of permissions. These apps may include newer or less visible apps that haven't yet been reviewed but still require broad access to device features. In benign apps, 4-star apps also showed high permission usage. These are typically well-rated, feature-rich applications, which may explain their need for more permissions to support a wide range of functions.

For malicious apps, we observed a general decline in permission requests over time, with fewer apps and lower permission usage across most star ratings. This aligns with previous observations that malicious apps may be scaling back permissions to avoid detection.

Conversely, benign apps showed a consistent increase in both app count and permission usage across all rating levels, especially in 0-star, 3-star, and 4-star apps. This likely reflects the growing complexity and functionality of newer apps and those maintaining moderate to high user engagement.

Overall, this trend highlights the need for users to remain cautious. Even highly rated apps can request extensive permissions, and 0-star apps, despite lacking visible user approval, often demand the most. Developers should remain transparent about why permissions are needed, and platforms should encourage best practices to prevent unnecessary or invasive access.

\begin{takeaway}
Benign apps increasingly request more permissions across all rating levels, especially at 0 and 4 stars. Malicious apps are trending toward fewer permissions, likely to reduce detection.
\end{takeaway}

\begin{table*}[t]
\centering
\caption{Number of malicious apps (A) and total permissions requested (P), categorized by the apps' star ratings and year.}\label{tab:maliciouspermissions_by_rating}
\vspace{-3mm}
\begin{tabular}{l|rr|rr|rr|rr|rr|rr}
\xl{2}
\multirow{2}{*}{Year} & 
\multicolumn{2}{c|}{\ding{73}\ding{73}\ding{73}\ding{73}\ding{73}} & 
\multicolumn{2}{c|}{\ding{72}\ding{73}\ding{73}\ding{73}\ding{73}} & 
\multicolumn{2}{c|}{\ding{72}\ding{73}\ding{73}\ding{73}\ding{73}} & 
\multicolumn{2}{c|}{\ding{72}\ding{72}\ding{72}\ding{73}\ding{73}} & 
\multicolumn{2}{c|}{\ding{72}\ding{72}\ding{72}\ding{72}\ding{73}} & 
\multicolumn{2}{c}{\ding{72}\ding{72}\ding{72}\ding{72}\ding{72}} \\
\cline{2-13}
& A & P & A & P & A & P & A & P & A & P & A & P \\
\xl{1}
2019 & 62 & 730 & 0 & 0 & 6 & 105 & 18 & 106 & 28 & 315 & 1 & 16 \\
2020 & 46 & 523 & 2 & 26 & 4 & 20 & 21 & 161 & 16 & 150 & 0 & 0 \\
2021 & 69 & 788 & 0 & 0 & 8 & 91 & 12 & 22 & 30 & 443 & 0 & 0 \\
2022 & 47 & 312 & 0 & 0 & 9 & 28 & 20 & 302 & 42 & 417 & 0 & 0 \\
2023 & 41 & 190 & 0 & 0 & 10 & 64 & 10 & 70 & 28 & 149 & 1 & 12 \\
\xl{2}
\end{tabular}
\vspace{-3mm}
\end{table*}

\begin{table*}[t]
\centering
\caption{Number of benign apps (A) and total permissions requested (P), categorized by the apps' star ratings and year.}
\label{tab:benign_permissions_by_star_ratings}
\vspace{-3mm}
\begin{tabular}{l|rr|rr|rr|rr|rr|rr}
\xl{2}
\multirow{2}{*}{Year} &
\multicolumn{2}{c|}{\ding{73}\ding{73}\ding{73}\ding{73}\ding{73}} & 
\multicolumn{2}{c|}{\ding{72}\ding{73}\ding{73}\ding{73}\ding{73}} & 
\multicolumn{2}{c|}{\ding{72}\ding{73}\ding{73}\ding{73}\ding{73}} & 
\multicolumn{2}{c|}{\ding{72}\ding{72}\ding{72}\ding{73}\ding{73}} & 
\multicolumn{2}{c|}{\ding{72}\ding{72}\ding{72}\ding{72}\ding{73}} & 
\multicolumn{2}{c}{\ding{72}\ding{72}\ding{72}\ding{72}\ding{72}} \\
\cline{2-13}
& A & P & A & P & A & P & A & P & A & P & A & P \\
\xl{1}
2019 & 315 & 1,646 & 4 & 11 & 27 & 228 & 82 & 884 & 214 & 2,655 & 8 & 74 \\
2020 & 414 & 5,687 & 4 & 23 & 23 & 261 & 110 & 1,435 & 173 & 2,436 & 7 & 51 \\
2021 & 485 & 6,238 & 4 & 83 & 38 & 442 & 139 & 1,848 & 225 & 3,251 & 5 & 40 \\
2022 & 523 & 6,237 & 7 & 160 & 68 & 858 & 199 & 2,869 & 303 & 4,666 & 1 & 3 \\
2023 & 505 & 5,459 & 10 & 98 & 54 & 560 & 196 & 2,732 & 278 & 4,200 & 8 & 113 \\
\xl{2}
\end{tabular}
\vspace{-3mm}
\end{table*}

\subsubsection{App Installs} App installs represent the number of times an application has been downloaded from the Google Play Store. Install count is often seen as an indicator of popularity or trust, but it does not always reflect the app's behavior in terms of permission usage.

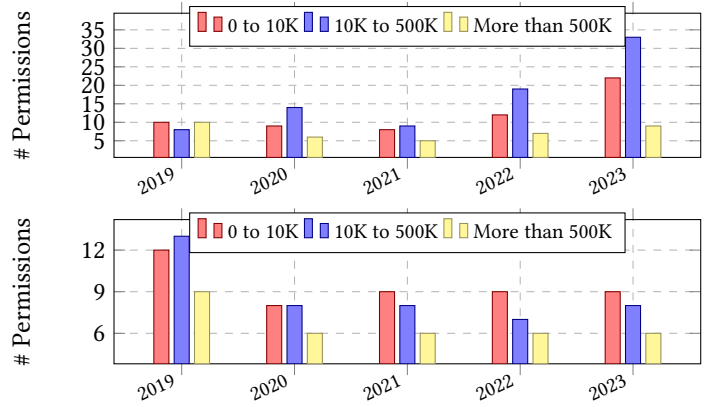
\begin{figure}[t]
\pgfplotsset{}
   \centering
\begin{tikzpicture}
    \begin{axis}
    [
        ybar,
        ymin = 5, ymax = 35,
        ytick distance = 5,
        enlargelimits=0.15,
        width = 5cm,
        height = 3.5cm,
        ylabel={\# Permissions},
        ylabel style ={font = \fontsize{10}{12}\selectfont},
        yticklabel style={font = \fontsize{10}{12}\selectfont},
        symbolic x coords={2019, 2020, 2021, 2022, 2023},
        x=1.5cm,
        bar width=2mm,
        legend style={at={(0.5,1.05)},anchor=north, legend columns=3, font=\fontsize{8}{10}\selectfont},
        xticklabel style={rotate=25,anchor=east, align=right, font=\fontsize{8}{10}\selectfont},
        grid=both,grid style={line width=0.1pt, draw=gray!10},major grid style={line width=.1pt,draw=black!30, dashed}
        ]       
    \addplot[ybar, fill=red!50!white, draw=red!50!black] coordinates {(2019,10) (2020,9) (2021,8) (2022,12) (2023,22)};
     \addlegendentry{0 to 10K};
    \addplot[ybar, fill=blue!50!white, draw=blue!50!black] coordinates {(2019,8) (2020,14) (2021,9) (2022,19) (2023,33)};
    \addlegendentry{10K to 500K};
    \addplot[ybar, fill=yellow!50!white, draw=yellow!50!black] coordinates {(2019,10) (2020,6) (2021,5) (2022,7) (2023,9)};
    \addlegendentry{More than 500K};    
    \end{axis}
\end{tikzpicture}

\pgfplotsset{}
   \centering
\begin{tikzpicture}
    \begin{axis}
    [
        ybar,
        ymin = 5, ymax = 13,
        ytick distance = 3,
        enlargelimits=0.15,
        width = 5cm,
        height =3.5cm,
        ylabel={\# Permissions},
        ylabel style ={font = \fontsize{10}{12}\selectfont},
        yticklabel style={font = \fontsize{10}{12}\selectfont},
        symbolic x coords={2019, 2020, 2021, 2022, 2023},
        x=1.5cm,
        bar width=2mm,
        legend style={at={(0.5,1.05)},anchor=north, legend columns=3, font=\fontsize{8}{10}\selectfont},
        xticklabel style={rotate=25,anchor=east, align=right, font=\fontsize{8}{10}\selectfont},
        grid=both,grid style={line width=0.1pt, draw=gray!10},major grid style={line width=.1pt,draw=black!30, dashed}
        ]       
    \addplot[ybar, fill=red!50!white, draw=red!50!black] coordinates {(2019,12) (2020,8) (2021,9) (2022,9) (2023,9)};
     \addlegendentry{0 to 10K};
    \addplot[ybar, fill=blue!50!white, draw=blue!50!black] coordinates {(2019,13) (2020,8) (2021,8) (2022,7) (2023,8)};
    \addlegendentry{10K to 500K};
    \addplot[ybar, fill=yellow!50!white, draw=yellow!50!black] coordinates {(2019,9) (2020,6) (2021,6) (2022,6) (2023,6)};
    \addlegendentry{More than 500K};    
    \end{axis}
\end{tikzpicture}\vspace{-3mm}
\caption{Requested permissions across download ranges for (top) malicious and (bottom) benign apps.}
\label{fig:Benign-Apps-Downloads}
\label{fig:Malicious-Apps-installs}
\vspace{-3mm}
\end{figure}

We analyzed how permission requests vary across different install ranges, focusing on both benign and malicious apps. For malicious apps, the 10K--500K download range consistently requested the highest number of permissions, far exceeding both lower and higher download groups, as shown in~\autoref{fig:Malicious-Apps-installs}. This suggests that mid-range apps may be particularly risky, as they appear popular enough to attract users but remain under the radar of stricter scrutiny. In contrast, high-download malicious apps (500K+) showed a notable decline in requests over time, possibly due to tighter enforcement or a strategic move to avoid raising red flags.

Benign apps exhibited more stable trends across all download ranges. While permissions in the 10K--500K range initially declined after 2019, the overall pattern remained consistent, as seen in~\autoref{fig:Benign-Apps-Downloads}. High-download benign apps showed the most controlled and predictable permission behavior, aligning with user expectations for trustworthy, well-established apps.

This comparison highlights an important privacy concern: apps with mid-level popularity, especially malicious ones, may exploit their visibility to gain excessive access. Users often assume that more downloads means more safety, but this is not always the case. Vigilance is needed not only for unknown apps, but also for those sitting in the middle tier of popularity.

\begin{takeaway}
Malicious apps in the 10K--500K download range request the most permissions, while high-download apps show more restraint. Benign apps are generally more consistent, but install count does not guarantee safe permission practices.
\end{takeaway}

\subsubsection{App Sizes} App size refers to the amount of storage space an application takes on a device, typically measured in megabytes (MB). We examined whether apps of different sizes request more permissions and what implications that has for privacy and security.

Our analysis compared benign and malicious apps across four size groups: very small (0--10 MB), small (10--25 MB), medium (25--50 MB), and large (50+ MB). We found that medium and large malicious apps consistently requested the most permissions overall, with a sharp spike in 2023. Notably, large malicious apps (50+ MB) jumped to 28 permissions on average that year, suggesting increased data access behavior among bigger malicious apps.

In contrast, benign apps demonstrated a stable and predictable pattern across all size categories. Medium-sized benign apps consistently requested the most permissions, large apps followed with moderate levels, and small or very small benign apps remained the most lightweight in terms of permission usage.

Very small and small malicious apps showed more erratic trends, including a noticeable drop in permissions in 2023 among very small apps. This inconsistency may point to varying strategies in how smaller malicious apps operate, sometimes attempting to appear less intrusive to avoid detection.

As shown in~\autoref{tab:permission_requests_onAPKSizes}, these findings suggest that malicious apps, particularly larger ones, are becoming more aggressive in permission use, possibly due to increased functionality or intent to access more sensitive data. Benign apps, on the other hand, continue to follow a more consistent and expected permission pattern.

\begin{takeaway}
Medium and large malicious apps request the most permissions, with a sharp rise in 2023. Benign apps remain stable, with predictable permission use based on app size.
\end{takeaway}

\begin{table}[t]
\centering
\caption{ Requests for Malicious (M) and Benign (B) Android Applications Categorized by App Size from 2019 to 2023.}\label{tab:permission_requests_onAPKSizes}\vspace{-3mm}
\scalebox{0.85}{
\begin{tabular}{l|cc|cc|cc|cc|cc}
\xl{2}
\hline
\multirow{2}{*}{App Size} & \multicolumn{2}{c|}{2019} & \multicolumn{2}{c|}{2020} & \multicolumn{2}{c|}{2021} & \multicolumn{2}{c|}{2022} & \multicolumn{2}{c}{2023} \\ \cline{2-11} 
                          & M & B & M & B & M & B & M & B & M & B \\ \hline
Very Small Apps 0-10 MB   & 6 & 8 & 10 & 6 & 5 & 5 & 8 & 8 & 5 & 4 \\ 
Small Apps 10-25 MB       & 10 & 9 & 9 & 8 & 6 & 8 & 9 & 7 & 8 & 9 \\ 
Medium Apps 25-50 MB      & 14 & 11 & 14 & 10 & 10 & 9 & 17 & 7 & 19 & 11 \\ 
Large Apps 50+ MB         & 9 & 9 & 7 & 7 & 7 & 8 & 12 & 7 & 28 & 7 \\ 
\hline
\xl{2}
\end{tabular}}\vspace{-3mm}
\end{table}


\section{Discussion}\label{sec:Discussion}

This study examines the Android permissions landscape, highlighting significant trends and patterns in permission requests across a diverse range of applications. The results reveal both malicious apps' evolving strategies and the increasing complexity and functionality of benign apps. This further underlines the dynamic nature of app development and security challenges on the Android platform.

\BfPara{More Requests, More Challenges} Users who download an app from the Play Store see two screens. The first provides information such as the app's description, reviews, and screenshots. The user must select ``Install'' to proceed to the next screen. The second displays the application's permissions in a clear, organized format. Installing the application grants all requested permissions automatically. These permissions are categorized to indicate their functionality, potential security implications, and privacy risks. For example, permissions related to location services like {\tt ACCESS\_FINE\_LOCATION} and {\tt ACCESS\_COARSE\_LOCATION} are grouped together.

Users can find detailed information about the permission by clicking or tapping on it. This helps them understand the potential risks of installing the application. For example, the {\tt READ\_CONTACTS} permission includes the description: ``Allows the app to read data about your contacts stored on your device, including the frequency with which you have called, emailed, or communicated in other ways with specific individuals''. The significant number of requested permissions highlights the possibility that many users might ignore examining those requested permissions and justification~\cite{FeltHEHCW12, RamachandranDGT17}, as done with other domains (e.g., web). 

\BfPara{Malware Requests} Android's support for addressing malware includes sandboxing each application and alerting users about the permissions requested by the app~\cite{MayrhoferSBK21, 0001B0SS15}. Each application operates as a separate process within its virtual machine. It does not have the permissions required to perform actions or access resources that could negatively impact the system or other apps. For instance, an application cannot make phone calls, access calendar events, or modify Wi-Fi settings by default. However, an app can explicitly request these privileges through permissions, and this study highlights a range of demonstrations of such requests. 

\BfPara{Less Permissions, More Maliciousness} One of the findings from our analysis is the notable reduction in permission requests by malicious apps over time. This trend suggests that developers likely adopt more sophisticated techniques to avoid detection \cite{BibiAMIMK20}, possibly by strategically minimizing their permission footprint. This stealthy strategy highlights the ongoing cat-and-mouse game between app developers and security researchers, where improvements in detection methodologies lead to more subtle evasion tactics by malicious actors. This underscores the need for continuous security mechanism advancements to counteract these evolving threats effectively. In contrast, benign apps have shown increased permission requests, particularly for sensitive permissions such as location, audio recording, and fingerprint use. This trend is likely driven by modern applications' growing complexity and feature sets, which require extensive permissions to deliver enhanced functionality. However, this raises privacy and security concerns, as users may grant access to sensitive data without fully understanding the implications. Our study emphasizes the importance of transparency and user education in permission management to mitigate the potential risks associated with excessive permission requests.

\BfPara{Ads and Privacy} The comparative studies conducted in this research offer deeper insights into how permissions are used across different app features and categories. For instance, apps that support advertisements or in-app purchases tend to request more permissions, which can be attributed to the need for these features to access various device functions and data. This finding highlights the need for careful monitoring and regulation of permission requests in ad-supported and commercial apps to protect user privacy. Additionally, our analysis revealed significant differences in permission requests across various app genres. For example, {\em finance} and {\em business} apps commonly request location and network permissions, while {\em educational} and {\em productivity} apps often require device management permissions. These insights provide a nuanced understanding of how different apps prioritize permissions based on their functionalities and user needs.

\BfPara{Dynamics and Oversight} The temporal scope of our study, from 2019 to 2023, allowed us to capture trends in permission requests over time. While permissions for malicious apps generally decreased, benign apps showed a more complex pattern with fluctuations in permission requests. This indicates that regulatory changes, user expectations, and technological advancements might be crucial to shaping permission usage practices. Our findings underscore the critical need for continuous monitoring, user education, and oversight to ensure user privacy and security in app permissions.

\BfPara{Implications and Future Integration} To preserve user trust, permission systems must become more transparent and aligned with user intent. Ethical app design means asking only for what's needed, and providing clear justifications. Our findings suggest that app store policies could be updated to incorporate automated tools that assess permission behavior, recommend safer alternatives, or flag suspicious patterns during the submission process. By integrating these insights into vetting systems, platforms like the Play Store can promote safer, more privacy-conscious app ecosystems.

\section{Limitations}\label{sec:Limitations}



\BfPara{AndroZoo} The dataset used was sourced from AndroZoo. Although we cross-referenced the apps with the Google Play Store to ensure they were available in the market, this approach still does not fully capture the current diversity and state of the store. Newly released or region-specific applications not included in our dataset might not reflected/represented in our findings, potentially affecting our results' generalizability.

\BfPara{Static Analysis} We rely on decompiled APKs to extract permissions, which do not account for dynamic permission requests during app runtime~\cite{WangWZWLLC22}. This approach might overlook certain permissions that apps request after installation, underestimating permission usage. Additionally, our study compares permissions and specific app features such as advertisements, in-app purchases, content ratings, and app sizes. While this provides valuable insights, it does not explore other potentially influential factors like user reviews, developer reputation, or app update frequency. These factors could also impact permission requests and app behavior.

\BfPara{VirusTotal} We leverage VirusTotal to distinguish between malicious and benign apps. However, this method may not capture the full spectrum of behaviors~\cite{Salem21}. The classification of apps as benign or malicious is based on available datasets and might not reflect the nuanced behaviors that fall between these categories. This binary classification could simplify app behaviors' complexity and associated risks. Moreover, while our study is extensive, its temporal scope spans from 2019 to 2023. It does not capture the early years of the Android ecosystem or the latest trends emerging post-2023. Rapid changes in app development practices could introduce new patterns in permission requests that our study does not capture.

\BfPara{Categorization} While we categorize permissions into higher-level semantic groups for structured analysis, this approach may overlook the unique implications of specific permissions~\cite{ZhauniarovichG16}. Grouping permissions can simplify analysis but obscure individual permissions' distinct risks and functionalities. Future research addressing these limitations could provide a more nuanced understanding of the Android permissions landscape. This could help refine privacy and security practices in the mobile app ecosystem.

\subsection{Recommendations} \label{sec:SafetyMeasurments}

In today's digital age, safeguarding privacy and security on mobile devices has become increasingly imperative. Android users, in particular, should protect their personal information from potential threats. To effectively enhance their privacy and security, users can adopt a series of proactive steps when dealing with app permissions on their Android devices: \begin{enumerate*}

\item \textbf{Review Permissions Before Installing Apps:} Users should check apps permissions before downloading them. Be cautious of apps that ask for excessive or unnecessary permissions.

\item \textbf{Use App Store Reviews and Ratings:} Users should look at reviews and ratings on official app stores. Apps with a high number of negative reviews mentioning privacy concerns or suspicious behavior should be avoided.


\item \textbf{Regularly Review App Permissions:} Users should periodically check the permissions granted to installed apps and revoke any that seem unnecessary. Android settings allow users to manage and review app permissions.

\item \textbf{Use Security Software:} Users should install reputable mobile security apps that detect and alert about potential threats and suspicious apps.

\item \textbf{Enable Google Play Protect:} Users should enable this built-in feature to scan their device and apps for harmful behavior, helping to keep the device secure.

\end{enumerate*}

\subsubsection*{Identifying Malicious Apps Based on Permission Requests}

Our analysis highlights several distinct trends that can aid in malware detection based on permission request patterns. These findings can also benefit users, enhancing their awareness through easily interpretable features that can be communicated to them.
\begin{enumerate*}

\item \textbf{Excessive Permissions:} Users should be wary of apps that request a large number of permissions, especially those that seem irrelevant to the app's core functionality (e.g., a flashlight app requesting access to your contacts).

\item \textbf{Permissions for Sensitive Data:} Users should be cautious of apps that ask for sensitive data access, such as camera, microphone, location, and contacts, without a clear need. This should raise a red flag.

\item \textbf{Unusual Combinations of Permissions:} Users should be cautious of apps requesting combinations of permissions that could compromise privacy (access to contacts and messaging).

\item \textbf{Frequent Updates/New Permissions:} Users should be aware app frequently updated where each update requests updated permissions might be a sign of malicious intent.

\end{enumerate*}

\subsubsection*{Measures to Protect Privacy on Android}

Protecting your privacy on Android devices is essential in an increasingly connected world. By taking proactive steps, you can safeguard your personal information and ensure a safer user experience. Here are some practical measures to enhance your Android privacy: 
\begin{enumerate*}

\item \textbf{Limit Data Sharing:} Users should be selective about sharing personal information with apps. They should use the app's settings to control what data it can access.

\item \textbf{Use Privacy-Focused Apps:} Users should opt for apps known for commitment to user privacy. They should look for apps with clear privacy policies and minimal permission requirements.

\item \textbf{Enable Two-Factor Authentication:} Users should enable two-factor authentication (2FA) for apps and services that support it, adding an extra layer of security to their accounts.

\item \textbf{Regular Software Updates:} Users should keep their operating system and apps updated to ensure they have the latest security patches and features.

\item \textbf{Monitor App Behavior:} Users should pay attention to how their apps behave. Sudden spikes in data usage, battery drain, or unusual behavior might indicate malicious activity.

\end{enumerate*}

\section{Conclusion and Future Work}\label{sec:Conclusion}
This study provides an analysis of the Android permissions landscape across app genres and time, for benign and malicious apps. Our findings reveal that malicious apps typically request fewer permissions, likely as a strategy to evade detection, while benign apps request more diverse and larger permission sets to support enhanced functionality. Through the application of the FP-Growth algorithm, we uncovered frequent permission combinations and co-occurrence patterns, providing deeper insights into the behaviors of both benign and malicious apps. This analysis, conducted across the entire dataset, over five years (2019--2023), and within 16 app genres, highlighted distinct permission usage patterns, such as the more targeted and minimalistic combinations in malicious apps and the higher diversity in benign ones. Ad-supported and in-app purchase-enabled apps were found to request more permissions, raising ongoing privacy concerns. Over time, apps exhibited some privacy-conscious development trends; however, risky and excessive permissions persist, particularly in specific genres like communication and gaming. These findings emphasize the critical need for user vigilance, developer transparency, and regulatory enforcement to mitigate permission misuse. By identifying both individual and combined permission request behaviors, this research offers actionable insights to improve app development practices, enhance user education, and foster safer digital ecosystems.

\BfPara{Future Work} Future research can build on this study by exploring several areas. Longitudinal studies extending the time frame could reveal long-term trends in permission requests and their effects on privacy and functionality. Real-time analysis of user behavior in response to permission requests may offer insights into how developers adapt their permission strategies. Advanced machine learning techniques could predict security risks based on permission patterns, leading to tools that detect over-permissioned or malicious apps. Comparative analyses between Android and other mobile operating systems could identify best practices for permission management, while research into user education programs on app permissions could empower users to make more informed privacy decisions. Lastly, integrating association rule mining approaches like FP-Growth into broader security frameworks could enhance app screening processes, offering proactive protection against permission misuse.


\end{document}